\documentclass[%
twocolumn,
 amsmath,amssymb,
 aps,authoryear
]{aastex63}
\usepackage{graphicx}
\usepackage{ulem}
\usepackage{epstopdf}
\usepackage{amsmath}
\usepackage{amssymb}
\usepackage{mathtools}
\usepackage{mathrsfs}
\usepackage{bm}
\setcitestyle{authoryear,round} 
\usepackage{color}
\usepackage{xcolor}
\usepackage{textcomp}
\usepackage{gensymb}
\usepackage{hyperref}
\newcommand{\rs}{\text{R}_\odot}
\newcommand{\dash}{\text{ -- }}
\newcommand{\lfoc}{\mathscr{L}}
\newcommand{\dperp}{D_\perp}
\newcommand{\dslab}{D_\perp^\text{slab}}
\newcommand{\dtwo}{D_\perp^\text{2D}}
\newcommand{\dslabsup}{D_\perp^{\text{slab}\ast}}
\newcommand{\btwo}{b_\text{2D}}
\newcommand{\bslab}{b_\text{slab}}
\newcommand{\DelR}{\langle \Delta R^2 \rangle}
\newcommand{\DelTheta}{\langle \Delta \theta^2 \rangle}
\newcommand{\bb}{\bm{B}}
\newcommand{\vb}{\bm{v}}

\newcommand{\Rb}{\bm{\mathcal{R}}}

\begin{document}
\title{Random Walk and Trapping of Interplanetary Magnetic Field Lines: Global Simulation, Magnetic Connectivity, and Implications for Solar Energetic Particles}
\author[0000-0002-7174-6948]{Rohit Chhiber}
\email{rohit.chhiber@nasa.gov}
\affiliation{Department of Physics and Astronomy, University of Delaware, Newark, DE 19716, USA}
\affiliation{Heliophysics Science Division, NASA Goddard Space Flight Center, Greenbelt MD 20771, USA} 
\author[0000-0003-3414-9666]{David Ruffolo}
\correspondingauthor{David Ruffolo}
\email{david.ruf@mahidol.ac.th}
\affiliation{Department of Physics, Faculty of Science, Mahidol University, Bangkok 10400, Thailand}
\author[0000-0001-7224-6024]{William H.~Matthaeus}
\affiliation{Department of Physics and Astronomy, University of Delaware, Newark, DE 19716, USA}
\affiliation{Bartol Research Institute, University of Delaware, Newark, DE 19716, USA}
%
\author[0000-0002-0209-152X]{Arcadi V.~Usmanov}
\affiliation{Department of Physics and Astronomy, University of Delaware, Newark, DE 19716, USA}
\affiliation{Heliophysics Science Division, NASA Goddard Space Flight Center, Greenbelt MD 20771, USA} 
%
\author{Paisan Tooprakai}
\affiliation{Department of Physics, Faculty of Science, Chulalongkorn University, Bangkok 10330, Thailand}
%
\author{Piyanate Chuychai}
\affiliation{33/5 Moo 16, Tambon Bandu, Muang District, Chiang Rai 57100, Thailand}
%
\author[0000-0002-5317-988X]{Melvyn L.~Goldstein}
\affiliation{University of Maryland Baltimore County, Baltimore, MD 21250, USA}
%
\begin{abstract}
The random walk of magnetic field lines is an important ingredient in understanding how the connectivity of the magnetic field affects the spatial transport and diffusion of charged particles. As solar energetic particles (SEPs) propagate away from near-solar sources, 
they interact with the fluctuating magnetic field, which  
modifies their distributions. 
We develop a formalism in which 
the differential equation describing the field line random walk contains both effects 
due to localized magnetic displacements and 
a non-stochastic contribution from the large-scale expansion.
We 
use this formalism together with a global magnetohydrodynamic simulation of the inner-heliospheric solar wind, which includes
a turbulence transport model, to estimate the diffusive spreading of magnetic field lines that originate in  different regions of the solar atmosphere. We first use this model to quantify field line spreading at 1 au, starting from a localized solar source region, and find rms angular spreads of about 20\degree\ - 60\degree. In the second instance, we use the model to estimate the size of the source regions from which field lines observed at 1 au may have originated, thus quantifying the uncertainty in calculations of magnetic connectivity; the angular uncertainty is estimated to be about 20\degree. Finally, we 
estimate the filamentation distance, i.e., the heliocentric distance up to which field lines originating in magnetic islands can remain strongly trapped in filamentary structures.  
We emphasize the key role of slab-like fluctuations in the transition from filamentary to more diffusive transport at greater 
heliocentric distances.
\end{abstract}
\section{Introduction}
Magnetic field lines 
are a useful construct frequently employed  
 \citep{parker1979} to 
help in determining the connectivity between points of 
observation \citep[][and references within]{owens2013lrsp}.
Connectivity is then employed to
determine the heat flux, which, in turn, helps to determine patterns of
plasma flow and the propagation 
of energetic particles.
Connectivity
influences 
energy transport, with 
possible impacts on plasma composition, 
reconnection, and other building blocks of space plasma 
studies \citep{suess1993AdSpR,crooker2006SSR}.
Mappings of coronal and 
interplanetary field
lines are employed in studies of coronal 
heating and acceleration and various elements of 
space weather 
\citep{tsurutani1981JGR,toth2005JGR,cranmer2007ApJS,antiochos2011ApJ,lario2017ApJ}.
Due to the importance of field line transport in these diverse studies, it is 
important to assess the
precision to which these field line descriptions 
are known, or can be known. 
This is not only a question of numerical accuracy; fluctuations enter as well, given that 
the interplanetary field
admits structure over scales ranging 
from $\sim1$ au down to 
kinetic scales.
Therefore, field lines themselves include  
multiscale patterns.
Consequently, 
uncertainties will be introduced if
the magnetic field is not adequately resolved
in numerical calculations. 
The relevant underlying theory for quantifying 
the effect
of fluctuations on magnetic field lines
was developed by  
\cite{jokipii1966cosmic} and by 
\cite{jokipii1969ApJstochastic}, 
who split the magnetic field into
a uniform mean part
and fluctuations that were treated 
statistically.
This led to the 
theory of 
magnetic field line random walk (FLRW), 
a diffusion-like process in space.
Despite the fact that the formalism is well-known, 
the associated effect of fluctuations 
is not often included in quantitative 
discussions
of the field line connectivity that
links the photosphere to the heliosphere \citep[e.g.,][]{lario2017ApJ}.\footnote{\footnotesize{See, however, \cite{ruffolo2003ApJ,laitinen2013ApJ,tooprakai2016ApJ,
laitinen2016AA,laitinen2018JSpWC}.}}


Energetic charged particles, including solar energetic particles (SEPs) that comprise a major component of space weather effects on human activity, largely follow magnetic field lines \citep{minnie2009JGR}.
Because of the FLRW, individual field lines frequently deviate from the large-scale field.
The effect of the FLRW on the transport of energetic charged particles perpendicular to the large-scale field has long been recognized and was initially modeled as diffusive \citep{jokipii1966cosmic}.
It is now known that after initial ballistic spreading, the ensemble average perpendicular transport is subdiffusive \citep{urch1977APSS} followed by a regime of asymptotic diffusion in which particles have separated from their initial field lines \citep{qin2002ApJ}, a process that we will refer to as cross-field transport \citep[see also Figure 1 of][]{Ruffolo2008ApJ}.

%
%


Also missing from the particle-diffusion picture are 
additional factors, such as solar wind expansion and topological 
trapping of field lines, which, acting separately or together, 
may confound simple analyses of
lateral transport. 
For example, observations of sudden decreases and increases in the density of SEPs from impulsive solar flares, termed as ``dropouts'', are inconsistent with a diffusive process and have instead been interpreted in terms of the filamentation of magnetic field line connectivity to a narrow source region at the Sun \citep{mazur2000ApJ}.
Thus, the initial propagation of SEPs 
can be a non-diffusive process connected 
with the magnetic field's topological properties and the trapping 
of field lines \citep{tooprakai2016ApJ,laitinen2017ApJ}.
It is also clear that 
the expansion and/or 
meandering of the large-scale field itself
may be influential in ways that homogeneous diffusion 
cannot account for. 
These two effects may act in concert if particles
are trapped in a flux tube that happens itself to experience an anomalously large lateral displacement, which may contribute to the wide longitudinal distributions reported from multi-spacecraft observations of SEPs from some impulsive solar events \citep[see, e.g.,][]{wibberenz2006ApJ,wiedenbeck2013ApJ,droge2014JGR}.
It is clear that 
some of these more subtle details of the 
field-line meandering problem will inevitably 
enter into 
the physics of the observed SEP spreading,
keeping in mind that complications such as 
highly distributed sources, especially in gradual events, 
may be also 
be a major factor. 

In discussing the effect of the field line random walk (or meandering) on field line realizations and SEP propagation, it is important
to consider the distinction between 
the statistics of field line length
and the statistics of field line spreading. 
The former problem is associated with pathlengths of SEP propagation \citep{zhao2019ApJ,laitinen2019ApJ}, and is considered in a separate paper (Chhiber et al. 2021, submitted). When considering fluctuations about the Parker field, one will likely find some paths that are shortened, and others lengthened, relative to the unperturbed field. While one may debate the significance of the effects of field line meandering, the pathlength problem is clearly distinct from that of field line spreading, 
which is the focus of the present work.

Here we 
address this problem 
by reconsidering the field line random walk in the 
context of an expanding, but not necessarily 
symmetric, large-scale field, including a standard 
FLRW  approach to 
simultaneously account for 
unresolved random fluctuations and the
displacements that are implied. 
We also account for temporary trapping and suppression of the FLRW 
in filamentary flux-tube structures \citep{ruffolo2003ApJ,tooprakai2007GRL}, which may explain the so-called ``dropouts'' seen in SEP observations \citep{mazur2000ApJ,ruffolo2003ApJ,tooprakai2016ApJ}.
The resulting model is applied to a description of 
field line spreading in a global 
three-dimensional (3D) magnetohydrodynamic (MHD) simulation of the inner heliosphere.

\begin{figure*}
\gridline{\fig{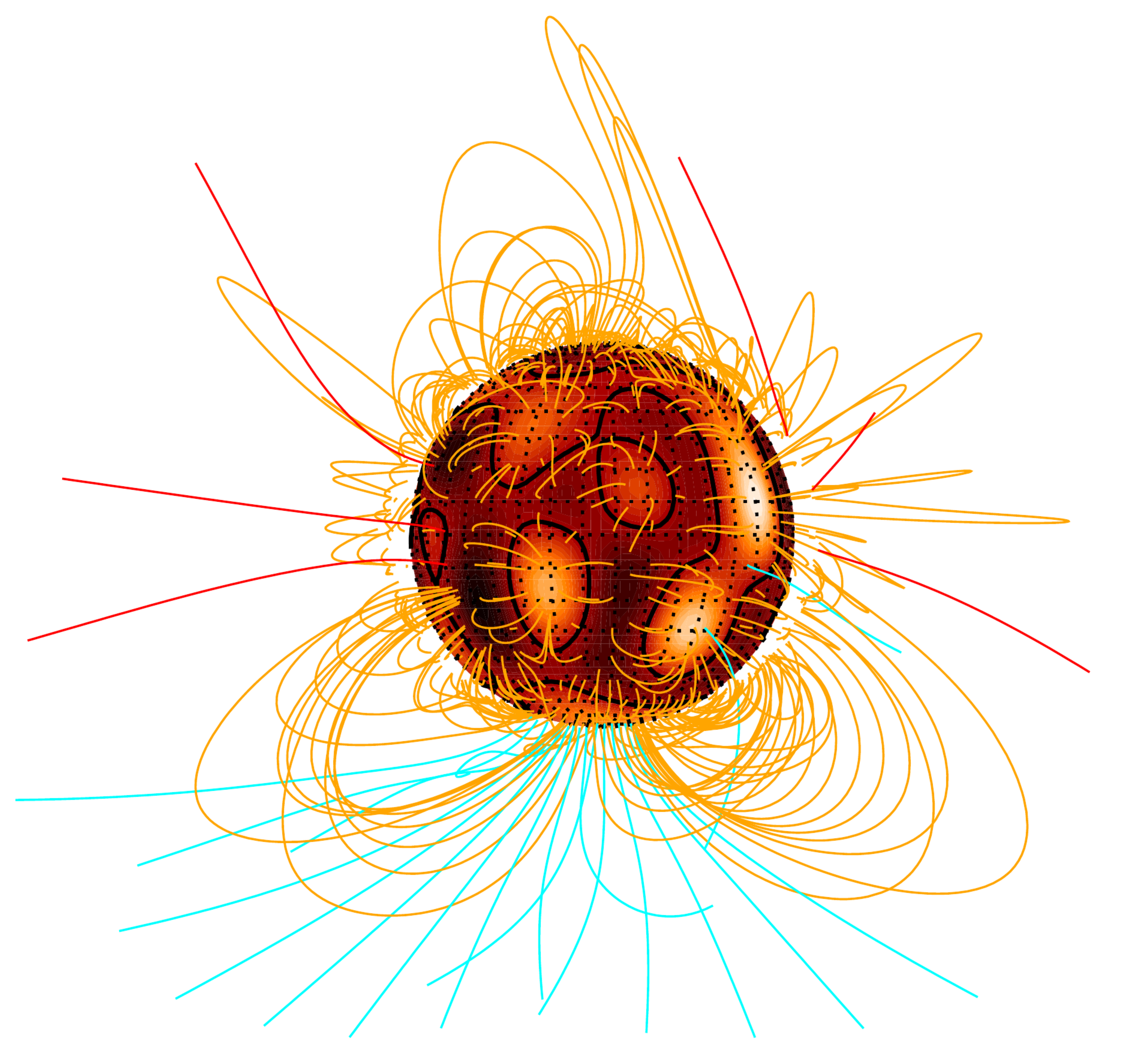}{.45\textwidth}{(a)}
         \fig{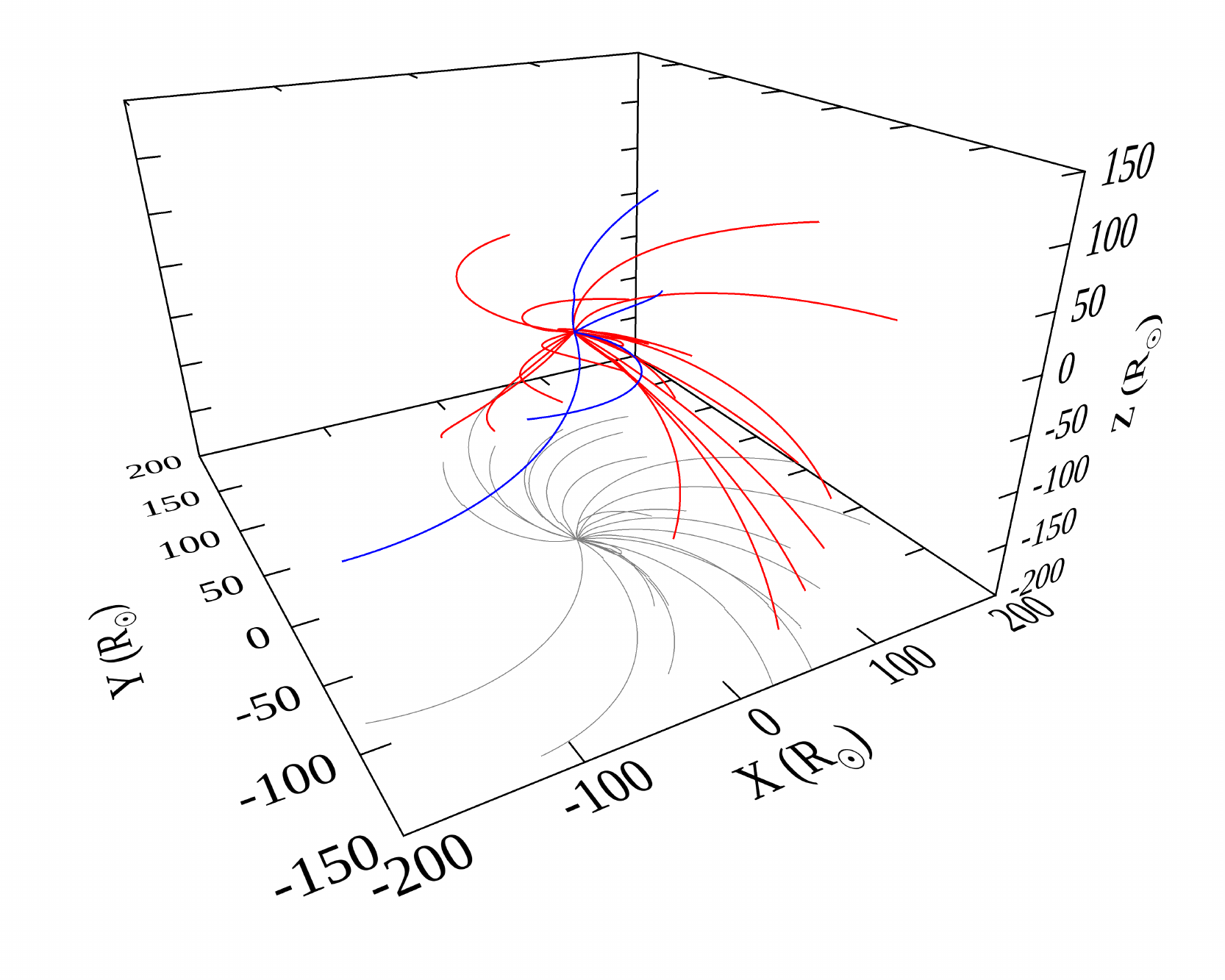}{.55\textwidth}{(b)}  
         }
\caption{Left and right panels show two views of magnetic field lines from Run II, based on a CR 2123 magnetogram (see Section \ref{sec:model_desc}). (a) A close-up view of the coronal base, with the center at 0\degree~heliolatitude and 116\degree~heliolongitude in heliographic coordinates \citep{franz2002pss}. Contours of radial magnetic field are shown at the coronal base. Yellow curves represent closed field lines.
(b) Open field lines traced out to 1 au.  Projections of field lines on the X-Y plane are shown in grey. In both panels blue and red curves represent open field lines with positive and negative polarity, respectively.}
\label{fig:hairy}
\end{figure*}
%

The elementary physics controlling field line transport is frequently discussed and its effects 
easily estimated.
Assuming diffusive transport in a non-expanding medium, and ignoring
complications such as temporary
topological trapping  \citep{ruffolo2003ApJ,chuychai2007ApJ},
the estimated 
root mean square (rms) displacement \(\delta R\)
of field lines transverse to the
mean magnetic field is 
\(\delta R \sim 2(Dz)^{1/2}\) 
for field line diffusion coefficient $D$ 
and 
displacement along the magnetic field $z$ \citep{matthaeus1995prl}.
If the ratio of the rms magnetic fluctuation to the mean magnetic field is of order unity ($\delta B/B_0 \sim 1$), then the field line diffusion
coefficient is generally of the order of a
turbulence outer scale $\lambda$ both for
quasilinear transport \citep{jokipii1966cosmic} and for 
nonlinear transport \citep{matthaeus1995prl}.
Accordingly, the transverse displacement at 1 au is  
roughly \(\delta R \approx 0.14 \text{--} 0.2\) au,
given that \(\lambda \approx 0.005 \text{--} 0.01\) au
near Earth orbit \citep{matthaeus2005,Ruiz2014SoPh}.
With this range of baseline estimates, 
the expected angular spread of field lines at 1 au is 
of the order of \(\sim 30\degree\), which is significant. 
This crude estimate takes into account 
neither expansion nor the spatial variation of
the magnetic field and its fluctuations.
In this work we develop an approach that 
includes these effects.

To set the stage for our study, in Figure \ref{fig:hairy} we show two views of magnetic field lines from our global MHD solar wind simulation (details in Section \ref{sec:sw_model}), using a Carrington Rotation (CR) 2123 magnetogram as an inner boundary condition, corresponding to a period of maximum solar activity. The smooth large-scale field lines shown here do not represent a complete picture, since the actual field lines may be significantly more complex 
due to unresolved magnetic turbulence 
\citep{matthaeus1995prl,matthaeus2003nonlinear,servidio2014ApJ,tooprakai2016ApJ}.
Given that broadband
Kolmogorov-like magnetic fluctuations are ubiquitous 
in the solar wind \citep{tu1995SSRv,matthaeus2011SSR,bruno2013LRSP,forman2013AIPC},
this effect of turbulence seems unavoidable. 
While  computational constraints do not permit the explicit representation of such fluctuations in a global solar wind code \citep{miesch2015SSR194,schmidt2015LRCA}, the statistical turbulence transport model coupled to our large-scale solar wind model allows us to estimate the average spread of field lines due to their random walk. We will use this approach to estimate both the spatial uncertainty in the source region of a field line at the coronal base [Figure \ref{fig:hairy}(a)] 
as well as the rms spread of meandering field lines near Earth [Figure \ref{fig:hairy}(b)].

In the following, Section \ref{sec:expan} presents an approach for quantifying effects of expansion on field lines. Section \ref{sec:flrw} briefly reviews the widely used two-component model, a convenient parameterization of magnetic fluctuations that facilitates manipulations 
describing turbulence. The same section adds into the field line transport model additional FLRW due to unresolved 
fluctuations. Section \ref{sec:filam} defines the filamentation distance, a measure of the heliospheric distance up to which field lines are strongly trapped in filamentary structures, and demonstrates that slab-like fluctuations play a key role in the escape of field lines from trapping based on the two-dimensional (2D) turbulence structure. 
We then present an application of this formalism to a 3D heliospheric simulation that includes turbulence transport. 
Section \ref{sec:sw_model} describes the simulation, while Section \ref{sec:results} presents results on mapping the diffusive spread of field lines from the lower solar atmosphere out to 1 au, and also treats the opposite case in which field lines localized at 1 au are mapped inward to the solar surface, or, more precisely, to the inner boundary of the simulation at the coronal base. 
Two types of simulation are considered: one with a dipole magnetic source and another employing a magnetogram. Finally, Section \ref{sec:disc} presents a summary and discussion of expected implications and applications of these results. 
\section{Development of Formalism}\label{sec:formal}
In this section we develop a 
formalism to describe the transport of magnetic field lines in an expanding medium. We consider three aspects of the problem: 1) the contribution of large-scale expansion to field line spreading;
 2) the effect of FLRW due to unresolved fluctuations; and 3) the trapping of field lines in filamentary structures. We examine the spreading of a statistical ensemble of field lines, relative to a central, resolved, large-scale field line, which  
 can be obtained from a model of the  interplanetary magnetic field (IMF) such as the Parker spiral, potential field source surface, or a global MHD simulation \citep{owens2013lrsp}.
\subsection{Effect of Expansion on Field Line Spreading}\label{sec:expan}

Even in the absence of magnetic field fluctuations, the large-scale expansion of the solar wind causes field lines to spread apart. Note that within the context of global solar wind simulations of the type we employ here (see Section \ref{sec:model_desc}), the effect of this expansion on the resolved large-scale magnetic field lines is incorporated 
{\it ipso facto}
in a self-consistent manner. Let the \(\hat{\bm{x}}\) and \(\hat{\bm{y}}\) unit vectors represent two mutually perpendicular directions that are transverse to the direction of a large-scale field line in the \(\hat{\bm{z}}\) direction. Consider next a field line at a small displacement (\(x, y\)) from the central large-scale field line of interest. If \(B_0\) denotes the magnetic field strength of this resolved (large-scale) 
field line, then the 
displaced field line
with magnetic field components (\(b_x,b_y,B_0\)) in the (\(x,y,z\)) coordinate system 
is described locally along the field line of interest by the equations 
\begin{equation}
\frac{dx}{dz} = \frac{b_x (x,y)}{B_0} 
= \frac{1}{B_0}\frac{\partial b_x}{\partial x}x 
+ \frac{1}{B_0}\frac{\partial b_x}{\partial y}y, \label{eq:expx}
\end{equation}
and
\begin{equation}
\frac{dy}{dz} = \frac{b_y (x,y)}{B_0} 
= \frac{1}{B_0}\frac{\partial b_y}{\partial x}x 
+ \frac{1}{B_0}\frac{\partial b_y}{\partial y}y. \label{eq:expy}
\end{equation}
%
%
The second equality in Equations \eqref{eq:expx} and \eqref{eq:expy} follows from a Taylor expansion of \(b_{x,y}\) around \((x,y)=(0,0)\), and it is understood that the the partial derivatives \(\partial b_x/\partial x\) and \(\partial b_x/\partial x\) are evaluated at \(x=y=0\). 

Defining \(\Delta R^2 = x^2 + y^2\), we also have
\begin{equation}
\frac{d (\Delta R^2)}{dz} = 2x\frac{dx}{dz} + 2y\frac{dy}{dz}. \label{eq:Rexp1}
\end{equation}
Neglecting effects of non-axisymmetric expansion, which are likely to be evened out by the FLRW, we consider an axisymmetric population of small field-line displacements characterized by \(\langle xy \rangle = 0\) and \(\langle x^2 \rangle = \langle y^2 \rangle = \DelR/2\), where \(\langle \cdots \rangle\) denotes an ensemble average. From Equations \eqref{eq:expx}, \eqref{eq:expy}, and \eqref{eq:Rexp1}, we then obtain 
\begin{equation}
\begin{aligned}
\frac{d \DelR}{dz} &= 
\frac{2}{B_0} \frac{\partial b_x}{\partial x} \langle x^2 \rangle +
\frac{2}{B_0} \frac{\partial b_y}{\partial y} \langle y^2 \rangle  \\
&= \frac{\DelR}{B_0} \left( \frac{\partial b_x}{\partial x} + \frac{\partial b_y}{\partial y} \right) \\
&= -\frac{\DelR}{B_0} \frac{dB_0}{dz}, \\
\end{aligned}
\end{equation}
where we have made use of the solenoidality of the magnetic field in the last step. We can define a \textit{focusing length} \(\lfoc(z)\), with \(\lfoc^{-1} = -(1/B_0) dB_0/dz\) \citep[e.g.,][]{roelof1969}, so that 
\begin{equation}
\frac{d \DelR}{dz} = \frac{\DelR}{\lfoc(z)}, \label{eq:expan}
\end{equation}
in the absence of magnetic fluctuations.  
\subsection{Field Line Random Walk}\label{sec:flrw}
In the presence of magnetic fluctuations that are polarized 
transverse to the mean magnetic field direction \(\hat{\bm{z}}\), diffusive behavior of the field line can be described in terms of displacements in the \(x\) and \(y\) directions \citep{matthaeus1995prl}:
\begin{equation}
\langle \Delta x^2\rangle = \langle \Delta y^2\rangle = 2 \dperp \Delta z, \label{eq:diff0}
\end{equation}
where \(\dperp\) is the perpendicular diffusion coefficient and we have assumed axisymmetry about \(\hat{\bm{z}}\). 
Note that the specialization to transverse
polarizations does not limit the 
potential for wavevectors (or gradients) 
of the field fluctuations to lie in any direction.
On adding the contribution of (perpendicular) FLRW to Equation \eqref{eq:expan}, the mean squared spread of field lines \(\DelR\) evolves according to the following ordinary differential equation (ODE):
\begin{equation}
\frac{d \DelR}{dz} = \frac{\DelR}{\lfoc(z)} + 4 D_\perp(z), \label{eq:ODE1}
\end{equation}
which may be integrated along the field line coordinate \(z\), following the specification of  \(D_\perp\). Note that the \(4\dperp\) term on the rhs of Equation \eqref{eq:ODE1} arises from the combination of the \(x\) and \(y\) components (\(\DelR = \langle  \Delta x^2\rangle  + \langle \Delta y^2\rangle \)),  which contribute \(2\dperp\) each.
Note that in writing 
Equation (\ref{eq:ODE1})
as the sum of two independent parts, 
it is assumed that displacements due to
large scale (resolved) magnetic fields
are uncorrelated with displacements due to any unresolved stochastic fluctuations. 

As an instructive special case of Equation \eqref{eq:ODE1}, consider a radial large-scale magnetic field, with \(B_0 \propto r^{-2}\) to conserve magnetic flux, where \(r\) is the radial coordinate in a Sun-centered frame. Then \(r\) is also the coordinate along the large-scale field line, and we have \(\lfoc = r/2\). In terms of an angular mean-squared spread of field lines \(\DelTheta\), we have \(\DelR = r^2 \DelTheta\). In the absence of magnetic fluctuations, Equation \eqref{eq:ODE1} then becomes \(d (r^2 \DelTheta)/dr = 2r \DelTheta\), or \(d\DelTheta /dr = 0\). This restates the 
obvious property that 
radial magnetic field lines maintain a constant angular spread. With magnetic fluctuations, Equation \eqref{eq:ODE1} becomes (for the case of a radial magnetic field) \(d \DelTheta/dr = 4 D_\perp(r)/r^2\). 

Before proceeding further with the specification of \(D_\perp\) we briefly review the two-component composite model of magnetic fluctuations \citep{matthaeus1990JGR}. This model provides a useful parameterization of anisotropic heliospheric turbulence in a relatively simple mathematical form, and it has found application in several space physics problems, including turbulence transport \citep{Oughton2011JGRA116,Zank2017ApJ835},  cosmic ray scattering and propagation \citep{bieber1996dominant,shalchi2009,
wiengarten2016ApJ833,chhiber2017ApJS230}, and transport of magnetic field lines \citep{chuychai2007ApJ,ghilea2011ApJ}. 

The two-component model assumes that the magnetic field can be written as \(\bm{B}=\bm{B}_0+\bm{b}(x,y,z)\), where \(\bm{B}_0\) is a mean field in the \(\hat{\bm{z}}\) direction and the fluctuating field \(\bm{b}\) is transverse to \(\bm{B}_0\).  By construction, \(\langle \bm{b}\rangle=0\) and the fluctuation is split into a ``slab'' and a 2D component, as \(\bm{b}(x,y,z) = \bm{b}_\text{slab}(z) + \bm{b}_\text{2D}(x,y)\). 
By definition, the 
slab fluctuations depend only on the coordinate along the mean field while 2D fluctuations depend only on the transverse coordinates. In Fourier space, slab fluctuations have wavevectors parallel to the mean field and 2D fluctuations have wavevectors perpendicular to the mean field \citep{oughton2015philtran}. Although this slab+2D model is a kinematic model that simplifies analytical and computational work, its physical basis 
is motivated by disparate 
dynamical processes: slab fluctuations may be likened to parallel propagating  Alfv\'enic fluctuations \citep{belcher1971JGR}, while the 2D fluctuations can arise due to the dominantly perpendicular cascade observed in anisotropic MHD turbulence \citep{shebalin1983JPP,oughton1994JFM,goldreich1995ApJ}. Solar wind turbulence has been shown to be characterized 
by a dominant (\(\sim 80\%\)) 2D component with a minor slab component \citep{bieber1996dominant}.  
The relative energies in 2D and slab components are variable, depending for example on the solar wind speed \citep{dasso2005ApJ}.
However, it is important to 
emphasize that the slab+2D model
is not a dynamical representation, 
as shown, e.g., in  \citet{ghosh1998JGR}.

For the purpose of specifying the perpendicular FLRW diffusion coefficient in the present study, we use the results of \cite{ghilea2011ApJ}, who, building on previous work \citep{matthaeus1995prl,ruffolo2003ApJ,chuychai2007ApJ,seripienlert2010apj}, examined FLRW in the context of slab+2D turbulence. We briefly summarize some salient points discussed in \cite{ghilea2011ApJ} below and we refer the reader to that paper for a more detailed and rigorous presentation. In the following we use \(b^2\) as a shorthand for the fluctuation energy \(\langle b^2\rangle\), so that \(b\) refers to the root mean squared fluctuation strength \(\sqrt{\langle b^2\rangle}\). The slab and 2D components are assumed to be statistically independent, so that \(b^2 = \bslab^2 + \btwo^2\). Further, we assume statistical axisymmetry about the mean magnetic field direction \(\hat{\bm{z}}\), resulting in statistically identical fluctuations in the \(\hat{\bm{x}}\) and \(\hat{\bm{y}}\) directions.

In terms of order-of-magnitude relations we have, from Equation \eqref{eq:diff0}:
\begin{equation}
\dperp \sim \left\langle\left(\frac{dx}{dz}\right)^2\right\rangle \ell \sim \frac{\langle b_x^2\rangle}{B_0^2}\ell,
\label{eq:diff_simple}
\end{equation}
where \(\ell\) is a ``mean free distance'' along the \(z\)-direction \citep{ruffolo2004ApJ} and in the second step we have used the field line equation \(dx/b_x = dz/B_0\). To prescribe \(\ell\) we must consider the process that determines the decorrelation of \(b_x\). For slab fluctuations \(b_x\) is a function only of \(z\), and the decorrelation is determined by the 
path of the field line along \(z\), following the mean field \(B_0 \hat{\bm{z}}\). Therefore one expects that \(\ell\) is of the order of \(\lambda_\text{slab}\), the slab correlation scale, so that the slab contribution to the FLRW is given by the quasilinear expression \citep{jokipii1968PRL,jokipii1969ApJstochastic,
matthaeus1995prl,ghilea2011ApJ}
\begin{equation}
\dslab = \frac{1}{2}\frac{b^2_\text{slab}}{B_0^2}\lambda_\text{slab}, \label{eq:Dslab}
\end{equation}
where we have used the fact that \(b^2 = 2\langle b_x^2\rangle\) for axisymmetric turbulence.

On the other hand, 2D fluctuations decorrelate when the transverse displacement is of the order of a perpendicular length scale of the turbulence, say \(\ell_\perp\). 
In the present work, we employ the assumption of random ballistic decorrelation (RBD): before the fluctuating field decorrelates, field line trajectories are assumed to be straight (ballistic) in random directions, for the purpose of calculating \(\dtwo\).  
This is in analogy with the standard calculation of a collisional mean free path for a gas, assuming straight-line motion between collisions.  
In particular, we use the RBD/2D assumption of \citet{ghilea2011ApJ}.
Then, for Equation \eqref{eq:diff0}, we relate the parallel mean free distance $\ell$ to the perpendicular mean free distance $\ell_\perp$ by  $\ell=(B_0/\sqrt{\langle b_x^2\rangle_\text{2D}})\,\ell_\perp$.
Eliminating \(\ell\) by using Equation \eqref{eq:diff_simple}, we get:
\begin{equation}
    \dtwo \sim \frac{\sqrt{\btwo^2/2}}{B_0}
    \ell_\perp.
\end{equation}
%
From a detailed calculation based on Corrsin's hypothesis, \cite{ghilea2011ApJ} obtained 
\begin{equation}
\dtwo = \frac{\sqrt{\pi}}{2}\frac{b_\text{2D}}{B_0} \lambda_\text{c2}, \label{eq:D2d}
\end{equation} 
where \(\lambda_\text{c2}\) is the 2D correlation scale. 

The perpendicular diffusion coefficient is then given by the direct sum of the slab and 2D contributions \citep[see][]{ghilea2011ApJ}:
\begin{equation}
    \dperp = \dtwo + \dslab.
\end{equation}
In Section \ref{sec:results} we will compute this \(D_\perp\) using the turbulence transport model in our global simulation, and solve Equation \eqref{eq:ODE1} along large-scale field lines. Note that other models for the diffusion coefficient can also be used to integrate Equation \eqref{eq:ODE1}.
\subsection{Estimation of the Filamentation Distance}\label{sec:filam}
Observations of ``dropouts'' of SEPs from impulsive solar flares, 
i.e., repeated sudden, non-dispersive decreases and increases of energetic particle fluxes \citep{mazur2000ApJ,gosling2004JGR}, 
have been associated with the filamentary distribution of magnetic field line connectivity from the solar corona to Earth orbit. 
This interpretation has been supported by numerical simulations based on the photospheric random motion of footpoints of the IMF \citep{giacalone2000ApJ}, or topological trapping in structures of anisotropic turbulence in the interplanetary medium \citep{ruffolo2003ApJ,zimbardo2004JGR}. These effects may, of course, be related in some way \citep[see, e.g.,][]{giacalone2006ApJ}.

These dropouts exist in contrast to observations of extensive lateral diffusion of energetic particles \citep{mckibben2005AdSpR,wibberenz2006ApJ,wiedenbeck2013ApJ,droge2014JGR}.
The persistence of sharp filamentary structures does not appear to be compatible with the statistically homogeneous approach 
used in fully diffusive models of field line transport \citep[e.g.,][]{jokipii1966cosmic,matthaeus1995prl,maron2004prl}. Within the 
framework of the two-component model introduced in the previous section, \cite{ruffolo2003ApJ} proposed that some field lines could 
be temporarily trapped in the small-scale topological structures of 2D turbulence. Such field lines typically 
originate near an O-point in the 2D turbulent field, and thus can be trapped in islands of closed orbits (which become filamentary 
flux tubes around O-lines in three dimensions), until they can escape due to the slab component of magnetic turbulence. Such escape can be 
suppressed by strong 2D fields. \cite{chuychai2005ApJ,chuychai2007ApJ} explored these concepts of temporary trapping and suppressed 
diffusive escape from trapping regions, which can provide a physical explanation for the near-Earth dropout observations. 
Here, we define a \textit{filamentation distance}, 
a distance from the Sun up to 
which field lines originating in magnetic islands can remain strongly trapped in filamentary structures, and 
employ those results to estimate this distance. 

\begin{figure*}
\centering
\includegraphics[width=.7\textwidth]{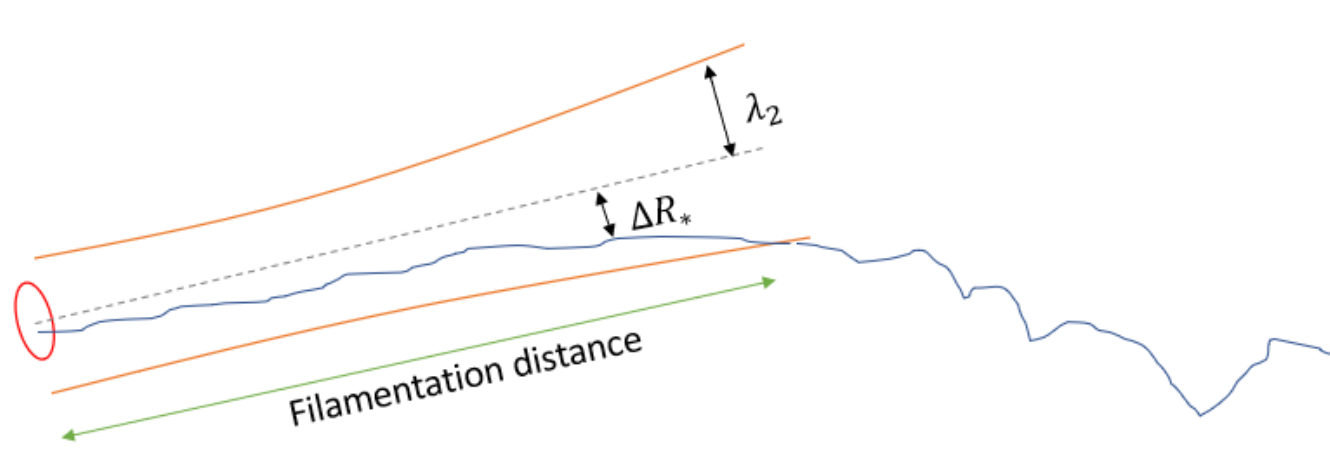}
\caption{Schematic of the \textit{filamentation distance} from the Sun, over which magnetic field lines can remain strongly trapped in filamentary flux tube structures. The solid blue curve represents a magnetic field line originating near an O-point in a magnetic island (red  oval) near the solar surface. The orange curves depict the boundary of a nominal magnetic flux tube surrounding the corresponding O-line (dashed line) with radius equal to the bendover scale of 2D turbulence \(\lambda_2\) (see text). The diffusion of the field line is suppressed over the filamentation distance, which is defined as the heliocentric distance at which the radial displacement \(\Delta R_\ast\) of the field line from the O-point (base of the dashed grey line) becomes larger than \(\lambda_2\). See also Figures 2 and 4 of \cite{chuychai2007ApJ}, which depict a similar physical process in numerical realizations of two-component magnetic turbulence.}
\label{fig:cartoon}
\end{figure*}

Using a quasi-linear approximation, \cite{chuychai2005ApJ} developed a theory to describe diffusion of field lines originating near an O-point of 2D turbulence.\footnote{\footnotesize{Such an O-point is a local maximum or minimum of the potential function of 2D magnetic fluctuations.}} The 2D fluctuations do not contribute to diffusive escape in such cases. However, there is a \textit{suppressed diffusive escape}, which may be thought of as an interference between the trapping 2D field and the escape-producing slab field: rapid circular motion within the trapping island leads to faster decorrelation of the slab turbulence observed by a random-walking field line. The suppressed slab field-line diffusion coefficient is given by \citep{chuychai2005ApJ}
\begin{eqnarray}
\dslabsup &=& \frac{\langle \Delta R_\ast^2\rangle}{2\Delta z} = \dslab \frac{P_{xx}^\text{slab}(K)}{P_{xx}^\text{slab}(0)}\\
&=& \dslab [1+(K l_z)^2]^{-5/6}, \label{eq:Dslab_sup}
\end{eqnarray}
where \(\langle \Delta R_\ast^2 \rangle\) is the mean-squared change in perpendicular distance \(R_\ast\) from an O-point of 2D turbulence, and we use \(K=\btwo/(B_0 \lambda_\text{c2})\) as a typical $z$-wavenumber for a field line to make a complete circuit of an island. In Equation (\ref{eq:Dslab_sup}), 
a particular form is adopted 
for the 
magnetic power spectrum 
of the \(x\)-component of the slab fluctuations;
in particular we use 
\(P_{xx}^\text{slab}(k_z) \propto [1 + (k_z l_z)^2)]^{-5/6}\), where 
\(k_z\) is a wavenumber parallel to the mean field and \(l_z\) is the bendover scale. This 
spectrum is flat for \(k_z \ll 1/l_z\), rolls over at \(k_z = l_z\), and has the Kolmogorov scaling \(k_z^{-5/3}\) for \(k_z \gg l_z\). For this spectrum, the bendover scale is related to the slab correlation scale by \(l_z = 1.339\lambda_\text{slab}\) \cite[see][]{chuychai2007ApJ}. 
As discussed in \cite{chuychai2005ApJ}, Equation \eqref{eq:Dslab_sup} tells us that the radial excursions\footnote{\footnotesize{That is, in the radial direction relative to the cylindrical coordinate system centered at the O-point (see 
Figure \ref{fig:cartoon}); not to be confused with the heliocentric radial direction.}} of field lines within such 2D islands is diffusive and is 
explicitly connected 
with the slab power spectrum evaluated 
at the wavenumber resonant with \(K\). Here \(K\) is the \(z\)-wavenumber corresponding to the initial unperturbed circular orbit within the island. For further details on the theory and verification by simulations we refer the reader to \cite{chuychai2005ApJ,chuychai2007ApJ}.

To estimate a filamentation distance, instead of Equation \eqref{eq:ODE1} we solve a different ODE for field lines initially trapped in small-scale topological structures, accounting for the associated suppression of slab diffusion described above.  
Considering a magnetic field line distribution with spread $\langle\Delta R_\ast^2\rangle^{1/2}$ around a radius $R_\ast\sim\langle\Delta R_\ast^2\rangle^{1/2}$, i.e., a distribution spread near the O-point, we combine expansion and diffusion terms to obtain
\begin{equation}
\frac{d \langle \Delta R_\ast^2 \rangle}{dz} = \frac{\langle \Delta R_\ast^2 \rangle}{\lfoc} + 2 \dslabsup, \label{eq:ODE2}
\end{equation}
where \(\dslabsup\) is given by Equation \eqref{eq:Dslab_sup}. 
We define the filamentation distance \(r_f\) as the heliocentric distance corresponding to \(z\) along the large-scale field line where \(\langle \Delta R_\ast^2 \rangle^{1/2} = \lambda_2\), the 2D bendover scale (see Figure \ref{fig:cartoon}), which is usually associated with the largest structures of 2D turbulence \citep{matthaeus2007spectral}.
Note that a smaller initial \(\langle \Delta R_\ast^2 \rangle^{1/2}\) implies that field lines start closer to the center of an O-point, and consequently experience stronger trapping \citep[as shown in][]{chuychai2007ApJ}. In Appendix \ref{sec:app} we consider a piecewise power-law 2D spectrum \citep{matthaeus2007spectral}, and show that the 2D correlation scale \(\lambda_{c2}\) can be related to the 2D bendover scale \(\lambda_2\) by \(\lambda_{c2}=\lambda_2(p+2)(\nu-1)/[(p+1)\nu]\), where \(p\) is the low-wavenumber power-law index and \(\nu\) is the inertial range power-law index. 
We choose \(p=2\), the lowest value consistent with  strict homogeneity of the two-point correlations, a 
property that may be questioned for the solar wind.
Note, however, that \cite{chhiber2017ApJS230} found that the value of \(p\) does not significantly influence energetic particle diffusion coefficients, so we adopt this value and defer further theoretical examination of this issue to a later time \citep[see also][and Appendix \ref{sec:app}]{engelbrecht2019apj}. 
With  \(p=2\) and \(\nu=5/3\) for a Kolmogorov inertial range, we obtain \(\lambda_{c2} = 0.53 \lambda_2\). This relation may be used to compute \(\lambda_2\) given \(\lambda_{c2}\) from the turbulence transport model we employ, described below in Section \ref{sec:model_desc}. 

\begin{figure*}
\centering
\includegraphics[width=.73\textwidth]{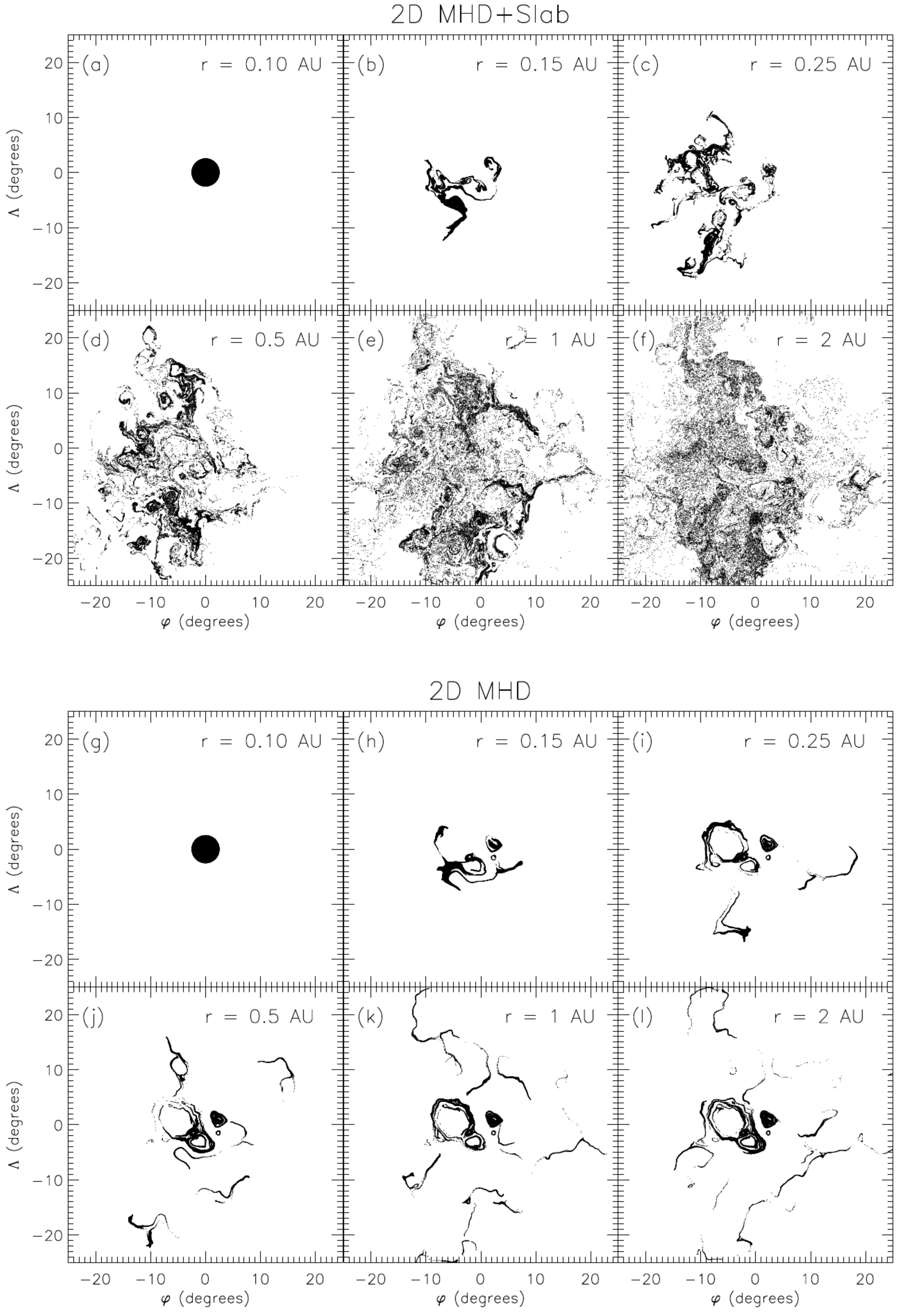}
\caption{(a)-(f): Trajectories of field lines in heliolongitude and heliolatitude as a function of radius \(r\) from the Sun, starting at random initial locations at r = 0.1 au within a circle of radius 2.5\degree, for a radial mean magnetic field superposed with a realization of 2D+slab magnetic turbulence, from \citet{tooprakai2016ApJ}. For this magnetic field, the calculated filamentation distance is 0.7 au, indicating a change from strong filamentation of field-line connectivity at \(r < 0.7\) au to a more diffusive spatial distribution, within a rather sharp outer envelope, at higher radius. \citet{tooprakai2016ApJ} showed that the energetic-particle distribution closely follows the field-line distribution, so this models the spatial distribution of SEPs from impulsive solar flares.  (g)-(l): Same, now removing the slab component of turbulence, i.e., for purely 2D MHD fluctuations. The difference in the field-line distribution confirms that slab fluctuations are crucial for defilamentation, i.e., for field lines to escape from 2D topological trapping and access wide regions within the overall angular spread of the FLRW.}
\label{fig:paisan}
\end{figure*}

To explain the significance of the filamentation distance, Figures \ref{fig:paisan}(a)-(f) 
show results from \cite{tooprakai2016ApJ}, of magnetic field line tracing for a radial mean field superposed with a realization of 2D+slab turbulence in which an initial random-phase 2D field was processed through a 2D MHD simulation. 
Here 50,000 field lines were traced from initial positions at a radial distance of 0.1 au from the Sun, randomly distributed in heliolongitude $\varphi$ and heliolatitude $\Lambda$ within a circle of angular radius 2.5\degree. The different panels show $\varphi$ and $\Lambda$ of the same set of 
magnetic field lines at different values of the radial distance $r$. This models the distribution of magnetic connections to a compact region near the Sun, such as the region within which SEPs are injected by an impulsive solar flare into the interplanetary medium. \cite{tooprakai2016ApJ} 
also confirmed by full orbital calculations that SEP trajectories closely follow the magnetic field line trajectories in this model \citep[see also][]{laitinen2013ApJ,laitinen2017ApJ}, and following previous reports \citep[e.g.,][]{giacalone2000ApJ,ruffolo2003ApJ,zimbardo2004JGR},
they associated the filamentary distribution of magnetic field-line connectivity with observations of ``dropouts'' of SEPs from impulsive solar flares.

For comparison, Figures \ref{fig:paisan}(g)-(l) show our new results of field line tracing from the same initial positions for the same 2D magnetic field, but now without slab fluctuations. 
In this case of pure 2D fluctuations superposed on the radial mean field, the field line coordinates $\varphi(r)$ and $\Lambda(r)$ are constrained to conserve the 2D magnetic potential $a(\varphi,\Lambda)$ \citep{tooprakai2016ApJ}. Thus, the angular coordinates $(\varphi,\Lambda)$ are constrained to remain on a single equipotential contour of $a$ \citep[such contours are plotted in Figure 1(c) of][]{tooprakai2016ApJ}. 
Level contours of 2D 
magnetic potential characteristically form closed 2D projections of flux tubes.
Consequently the presence of a strong 2D turbulence component provides 
a template for restricting field line excursions and cross field particle transport.
This strong constraint on FLRW  makes it 
impossible for field lines to wander in a purely 
diffusive manner, 
and 
also implies that field lines from nearby initial positions usually remain close together,
with strong concentrations of field line connectivity along narrow filaments and no connectivity with neighboring regions of space.
While 
many of the field lines are trapped along small equipotential contours that remain near the initial positions, there are some equipotential contours that do extend over significant angular distances (up to \(\sim25\degree)\). Thus, even without slab turbulence, as $r$ increases out to 1 au,
the rms angular spread of field lines continues to increase, but the slab turbulence component is crucial for defilamentation. 


For the magnetic field in this 2D+slab simulation, we calculate the filamentation distance 
to be $r_f=0.7$ au.  In Figures \ref{fig:paisan}(b)-(c), where $r\ll r_f$, there is strong topological trapping of field lines and strong filamentation within a rms angular spread that increases with $r$.
Then in Figures \ref{fig:paisan}(d)-(f) where $r\sim r_f$ or is somewhat larger than $r_f$, 
there is some access of field lines (and energetic particles) to most angular regions (in a coarse-grained sense) within an outer envelope, and it is this less strongly filamented pattern, at or beyond the filamentation distance, 
that is qualitatively consistent with observed dropouts near Earth. Nevertheless, the spatial distribution of magnetic connectivity, or ``dropout pattern,'' remains substantially filamented at $r=1$ au, and indeed observations of dropouts confirm
that 2D-like structures maintain substantial integrity to $r\approx1$ au. Therefore, the lateral transport of field lines and energetic particles moving outward from the Sun
to Earth orbit
is not well described as diffusive.
Furthermore, Figure \ref{fig:paisan} \citep[as well as Figure 5 of ][]{tooprakai2016ApJ} indicates a rather sharp outer boundary of the lateral transport that is inconsistent with a purely diffusive process.

Given the role of the slab component in disrupting
filamentation, and presumably eventually leading to lateral diffusion at 
$r\gg1$ au, we have defined the filamentation distance based on diffusion associated with the slab component. 
This slab diffusion is necessary to escape topological trapping near O-points of the 2D field, and is in turn suppressed by a strong 2D field, as described earlier in this section.
This is why we chose to define $r_f$ as the distance where the field line displacement due to suppressed slab diffusion reaches the bendover scale of the 2D turbulence.

\section{Solar Wind Model with Turbulence Transport}\label{sec:sw_model}
To evaluate the rms displacement of heliospheric magnetic field lines using the formalism developed in the previous Section, we must specify both the large-scale magnetic field and the field-line diffusion coefficient, the latter of which requires quantifying the  smaller-scale turbulence properties along the field line. We emphasize that the formalism of Section \ref{sec:formal} can be quite generally used with any model that provides the requisite inputs. In this section we briefly describe the MHD solar wind model that we employ here to supply the requisite information. We also provide details on the numerical implementation. Note that an earlier version of this model was recently used by \cite{chhiber2017ApJS230} to compute 3D distributions of energetic-particle diffusion coefficients throughout the inner heliosphere.
\subsection{Model Description}\label{sec:model_desc}

The two-fluid MHD coronal and heliospheric model that we employ 
is described in detail in \cite{usmanov2014three} and \cite{usmanov2018}. 
The model is based on the 
Reynolds-averaging approach \citep[e.g.,][]{mccomb1990physics}: a physical field, e.g., $\tilde{\bm{a}}$, is separated into a mean and a fluctuating component: \(\tilde{\bm{a}} = \bm{a}+\bm{a}'\), making use of an ensemble-averaging operation where \(\bm{a} = \langle \tilde{\bm{a}} \rangle\) and, by construction, \(\langle\bm{a}'\rangle=0\). Application of this Reynolds
decomposition to the underlying primitive compressible MHD equations,
along with a series of approximations appropriate to the solar
wind, leads us to a set of mean-flow equations that are coupled
to the small-scale fluctuations via another set of equations for
the statistical descriptors of the unresolved
turbulence.

To derive the mean-flow equations, the velocity and magnetic fields are Reynolds-decomposed: $\tilde{\bm{v}} = \bm{v}+\bm{v'}$ and $\widetilde{\bm{B}} = \bm{B}+\bm{B'}$, and then substituted into the momentum and induction equations in the frame of reference corotating with the Sun. The ensemble averaging operator $\langle\dots\rangle$ is then applied to these two equations \citep{usmanov2014three,usmanov2018}. The resulting mean-flow model consists of a single momentum equation and separate ion and electron temperature equations, in addition to an induction equation. 
Density fluctuations are neglected, and pressure
fluctuations are only those required to maintain incompressibility \citep{zank1992waves}. The Reynolds-averaging procedure introduces additional terms in the mean flow equations, 
representing the influence of turbulence on the mean (average) dynamics. These terms involve the Reynolds stress tensor \(\Rb = \langle\rho\vb'\vb' -
\bb'\bb'/4\pi\rangle\), the mean turbulent electric field
\(\mathcal{\bm{\varepsilon}}_m = \langle\vb'\times\bb'\rangle(4\pi\rho)^{-1/2}\), the
fluctuating magnetic pressure \(\langle B'^2\rangle/8\pi\), and the turbulent heating, or ``heat function'' \(Q_T(\bm{r})\), which 
is apportioned between protons and electrons. Here the mass density \(\rho=m_p N_S\) is defined in terms of the proton mass \(m_p\) and number density \(N_S\). The pressure equations employ the natural ideal gas 
value of 5/3 for the adiabatic index.
The pressure equations 
also include weak proton-electron collisional friction terms involving a classical Spitzer collision time scale
\(\tau_{SE}\) \citep{spitzer1965,hartle1968ApJ151}
to model the energy exchange between the
protons and electrons by Coulomb collisions
\citep[see][]{breech2009JGRA}. We neglect the heat flux carried by protons; the electron heat flux (\(\mathbf{q}_E\)) below \(5\text{ -- }10~\rs\) is approximated by the classical collision-dominated model of \cite{spitzer1953PhRv} \citep[see also][]{chhiber2016solar}, while above \(5 \text{ -- } 10~\rs\) we adopt Hollweg's ``collisionless'' 
conduction model \citep{hollweg1974JGR79,hollweg1976JGR}. See \cite{usmanov2018} for more details. 

Closure of the above system 
requires a model for unresolved turbulence.
Although the Reynolds decomposition is not formally a scale separation, we
have in mind that the stochastic components treated as
fluctuations reside mainly at the relatively small scales, i.e., comparable to, or smaller than, 
the correlation scales.
Transport equations for the fluctuations may be obtained by subtracting the mean-field equations from the full MHD equations and averaging the difference  \citep[see][]{usmanov2014three}. This yields a set of three equations \citep{breech2008turbulence,usmanov2014three,usmanov2018}
for the chosen statistical descriptors of turbulence, namely \(Z^2 = \langle v'^2 + b'^2\rangle\), i.e., 
twice the fluctuation energy
per unit mass where \(\bm{b}' = \bb'(4\pi\rho)^{-1/2}\);
the normalized cross helicity,
\(\sigma_c
= 2\langle\vb'\cdot\bm{b}'\rangle/Z^2\), 
or normalized cross-correlation between velocity and magnetic field fluctuations;
and $\lambda$, a correlation
length perpendicular to the mean magnetic field. 
Other parameters include the normalized energy difference, which we treat as a constant parameter (\(=-1/3\)) derived from
observations \citep[see also][]{Zank2017ApJ835}, and the K\'arm\'an-Taylor
constants \cite[see][]{matthaeus1996jpp,smith2001JGR,
breech2008turbulence}. 

Note that the fluctuation energy loss due to von K\'arm\'an decay \citep{karman1938prsl,hossain1995PhFl,
wan2012JFM697,wu2013prl,bandyopadhyay2018prx}  is balanced in a quasi-steady state by internal energy supply in the 
pressure equations. To evaluate the Reynolds stress we assume that the turbulence is transverse to the mean field and axisymmetric about it \citep{oughton2015philtran}, so that we obtain \(\Rb/\rho = K_R(\bm{I} - \hat{\bm{B}}\hat{\bm{B}})\), where \(K_R = \langle v'^2 - b'^2\rangle/2 = \sigma_D Z^2/2\) is the residual energy and \(\hat{\bm{B}}\) is a unit vector in the direction of \(\bm{B}\). The turbulent electric field is neglected here. For further details see \cite{usmanov2018}.

%

\subsection{Numerical Implementation}
We solve the Reynolds-averaged mean-flow equations concurrently with the turbulence transport equations in the spherical shell between the base of the solar corona (just above the transition region) and the heliocentric distance of 5 au. The computational domain is split into two regions: the inner (coronal) region of 
\(1\dash 30~\rs\) and the outer (solar wind) region from \(30~\rs\dash 5\) au. The relaxation method, i.e., the integration of time-dependent equations in time until a steady state is achieved, is used in both regions.

The model is well tested and provides reasonable agreement with both in-situ and remote observations \citep{usmanov2011solar,usmanov2012three,usmanov2014three,
chhiber2017ApJS230,chhiber2018apjl,
usmanov2018,chhiber2019psp1,bandyopadhyay2020ApJS_cascade,ruffolo2020ApJ}. The simulations used have typically been of two major types, distinguished by the
inner surface magnetic boundary condition: 
In the first type a Sun-centered 
dipole magnetic field is imposed at the inner boundary, with a specified 
tilt angle relative to the solar rotation axis. Zero or small tilt angle is often associated with solar activity minimum, while larger tilt angles are a suitable approximation for the more disordered heliosphere during solar maximum conditions \citep{owens2013lrsp}. 
The second 
 kind of inner magnetic boundary condition 
is based on suitably normalized magnetograms \citep{riley2014SoPh,usmanov2018}. 

In the present paper 
we employ two runs: 
A dipole-based run with a tilt of 10\degree~relative to the solar rotation axis, toward 330\degree~longitude (Run I), and a Carrington Rotation (CR) 2123 magnetogram-based run representative of solar maximum conditions (Run II), during the time period 2012 April 28 to 2012 May 25. The CR 2123 magnetogram from Wilcox Solar Observatory was scaled by a factor of 8 and smoothed using a spherical harmonic expansion up to 9th order. Run I has a resolution of \(702\times 120\times 240\) along \(r\times \theta\times \phi\) coordinates. The computational grid has logarithmic spacing along the radial (\(r\)) direction, with the grid spacing becoming larger as \(r\) increases. The latitudinal (\(\theta\)) and longitudinal (\(\phi\)) grids have equidistant spacing, with a resolution of 1.5\degree~each.  
Run II has a resolution of \(154\times 60\times 120\) in the inner (coronal) region, and a resolution of \(400\times 120\times 240\) in the outer region.\footnote{\footnotesize{We remark here that our primary interest lies in field lines close to the ecliptic region at 1 au, but the dipole-based run will only produce open field lines at high heliocentric latitudes. As we shall see, however, even high-latitude field lines may be displaced to the ecliptic due to the FLRW and expansion.}}

\begin{figure*}
\gridline{\fig{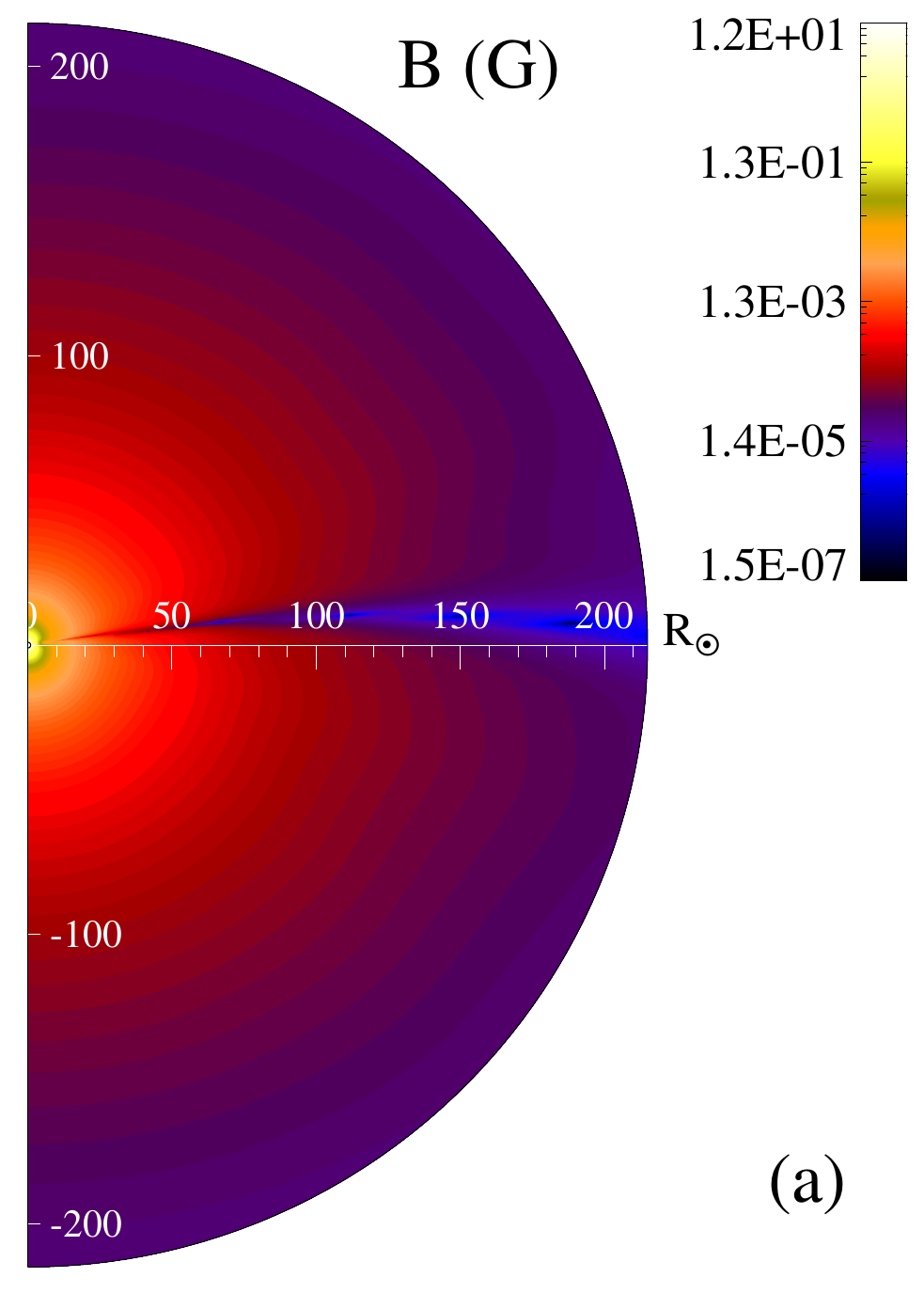}{.245\textwidth}{}
\fig{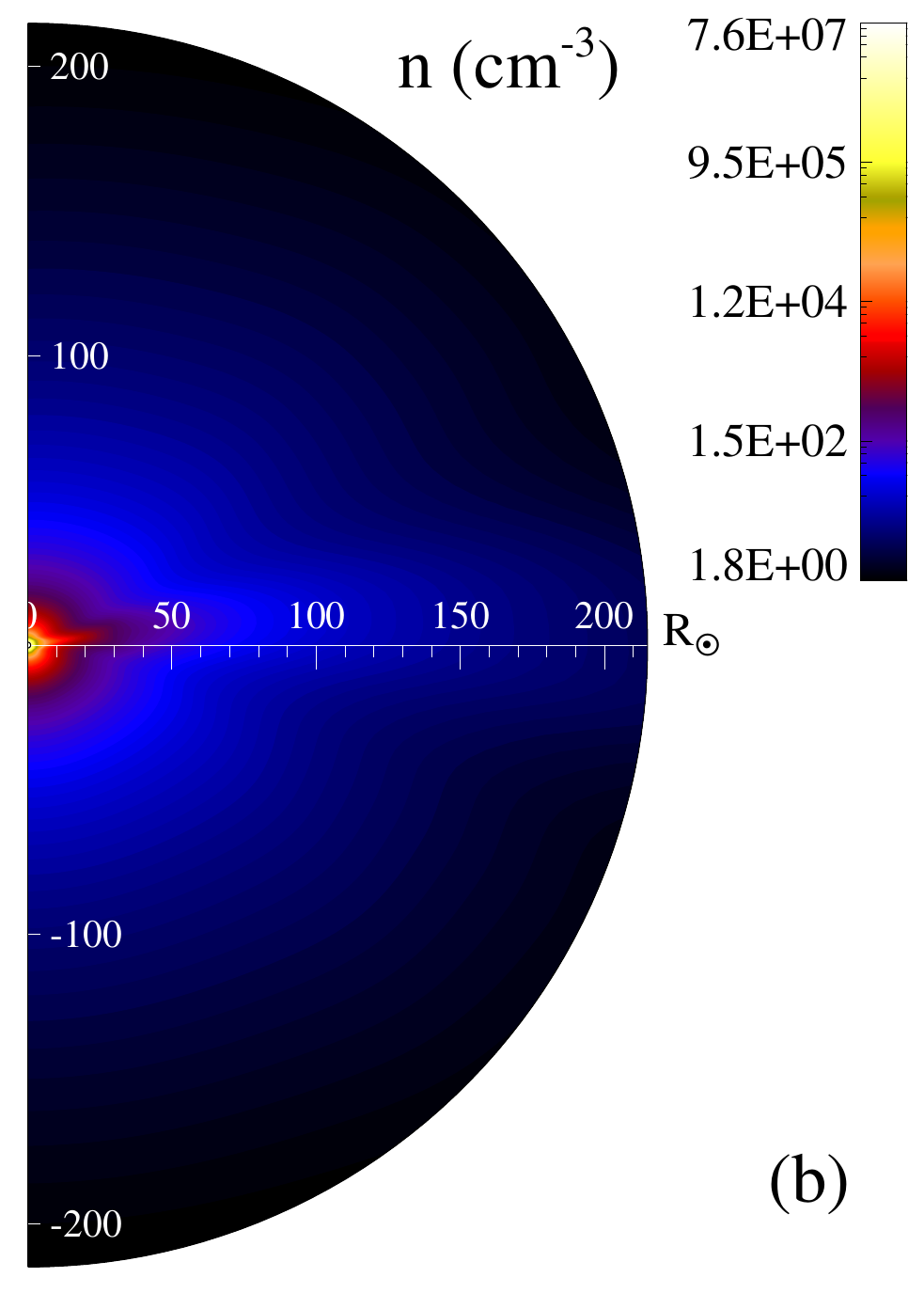}{.245\textwidth}{}
\fig{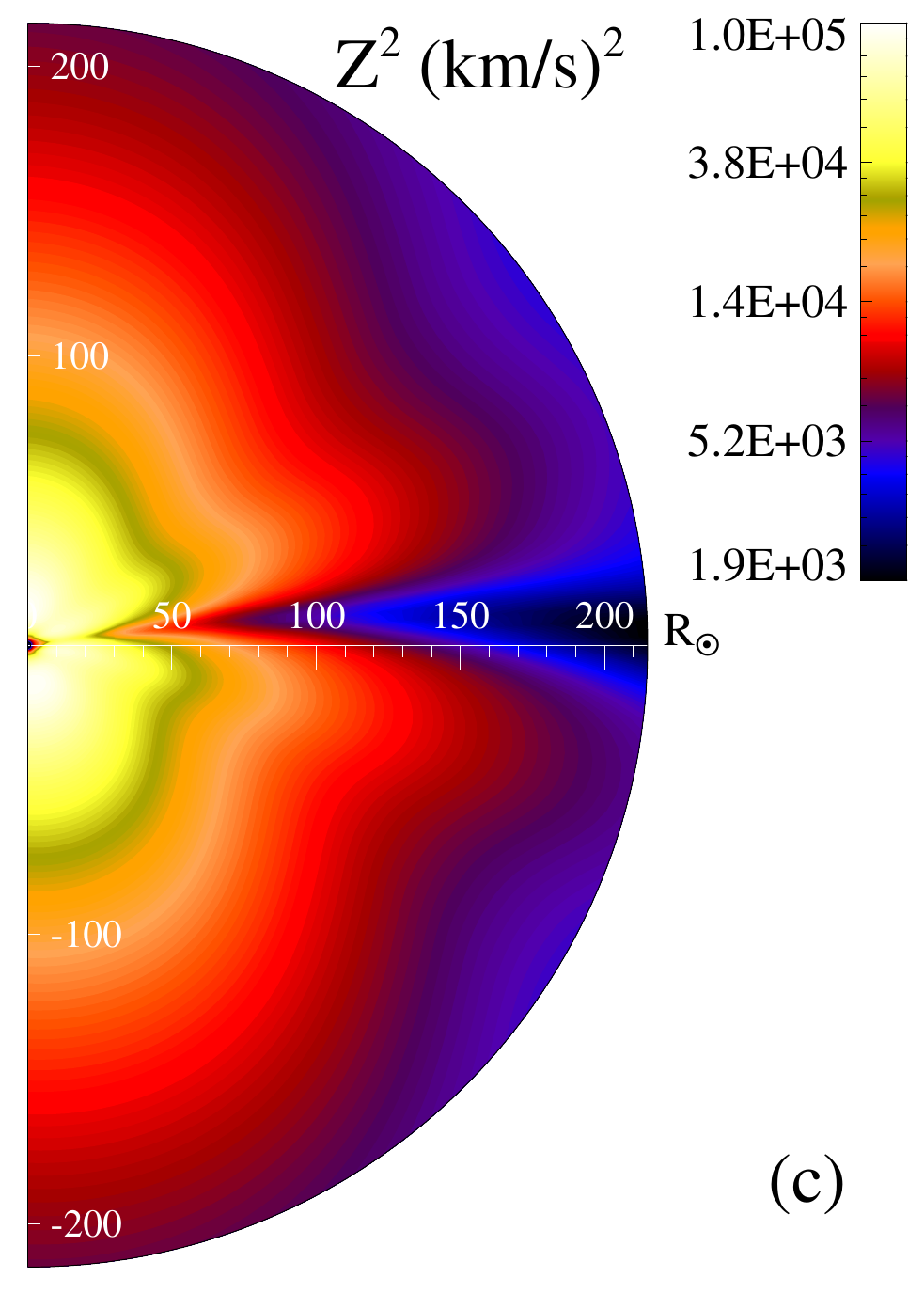}{.245\textwidth}{}
\fig{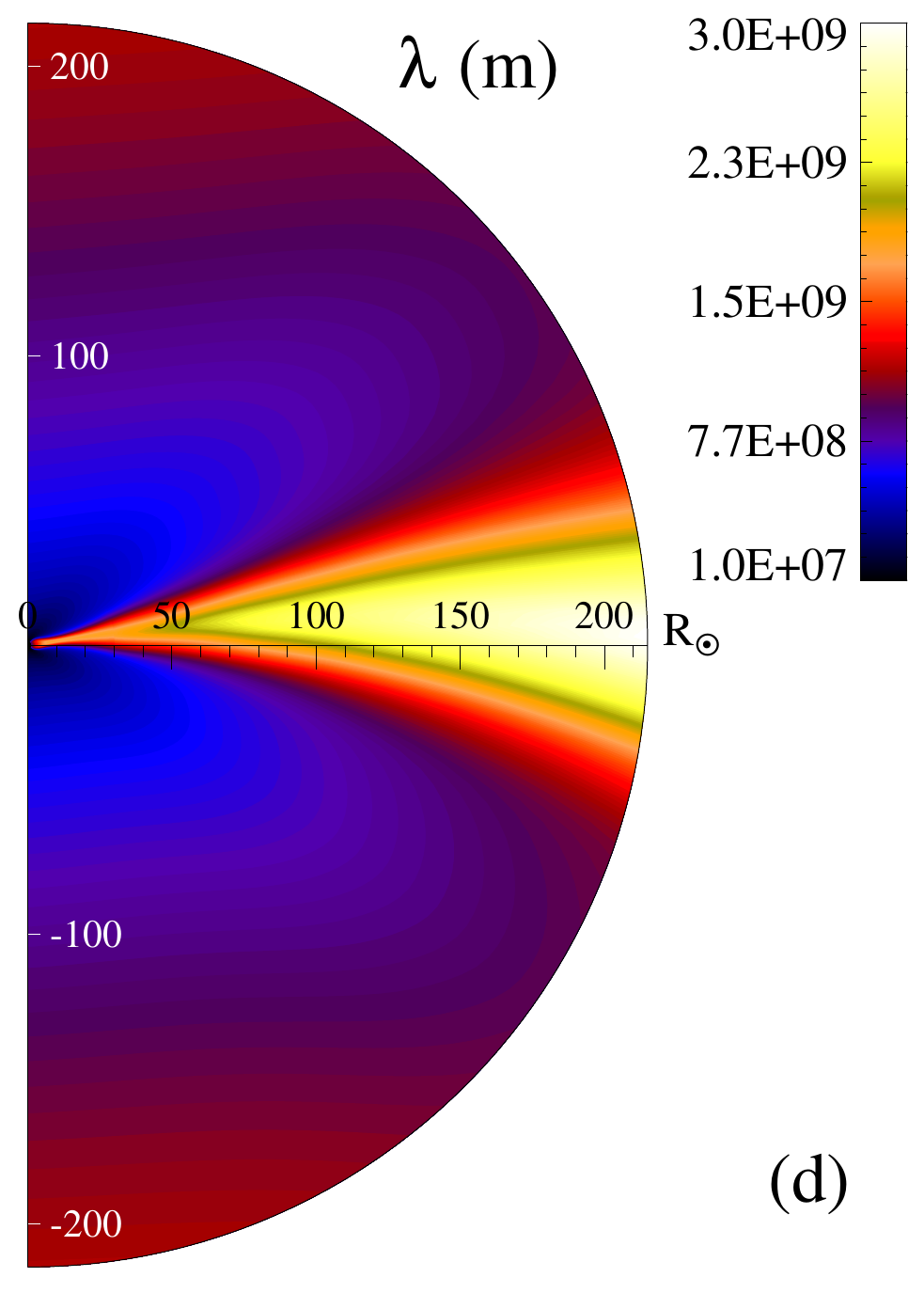}{.245\textwidth}{}}
\gridline{
\fig{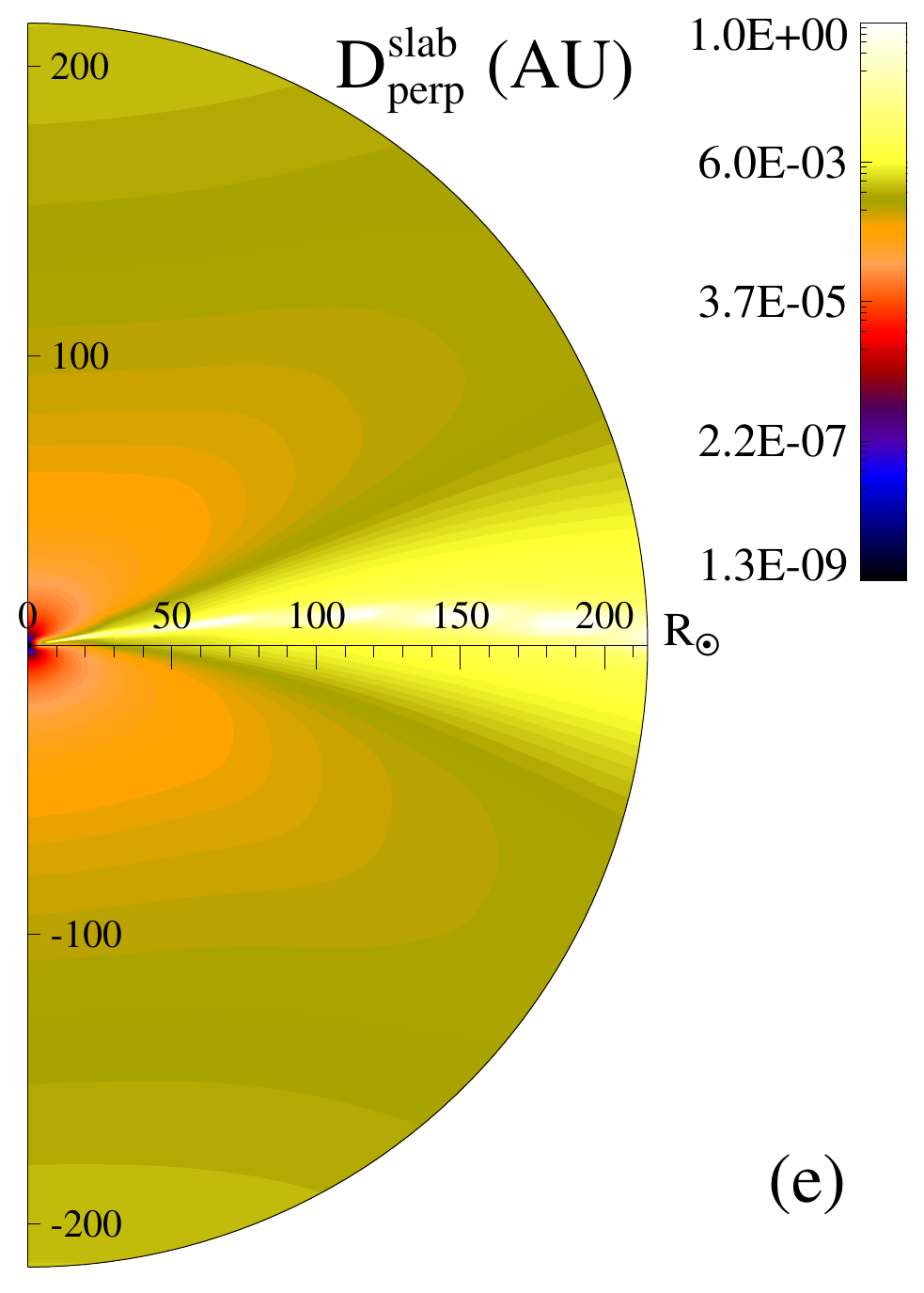}{.245\textwidth}{}
\fig{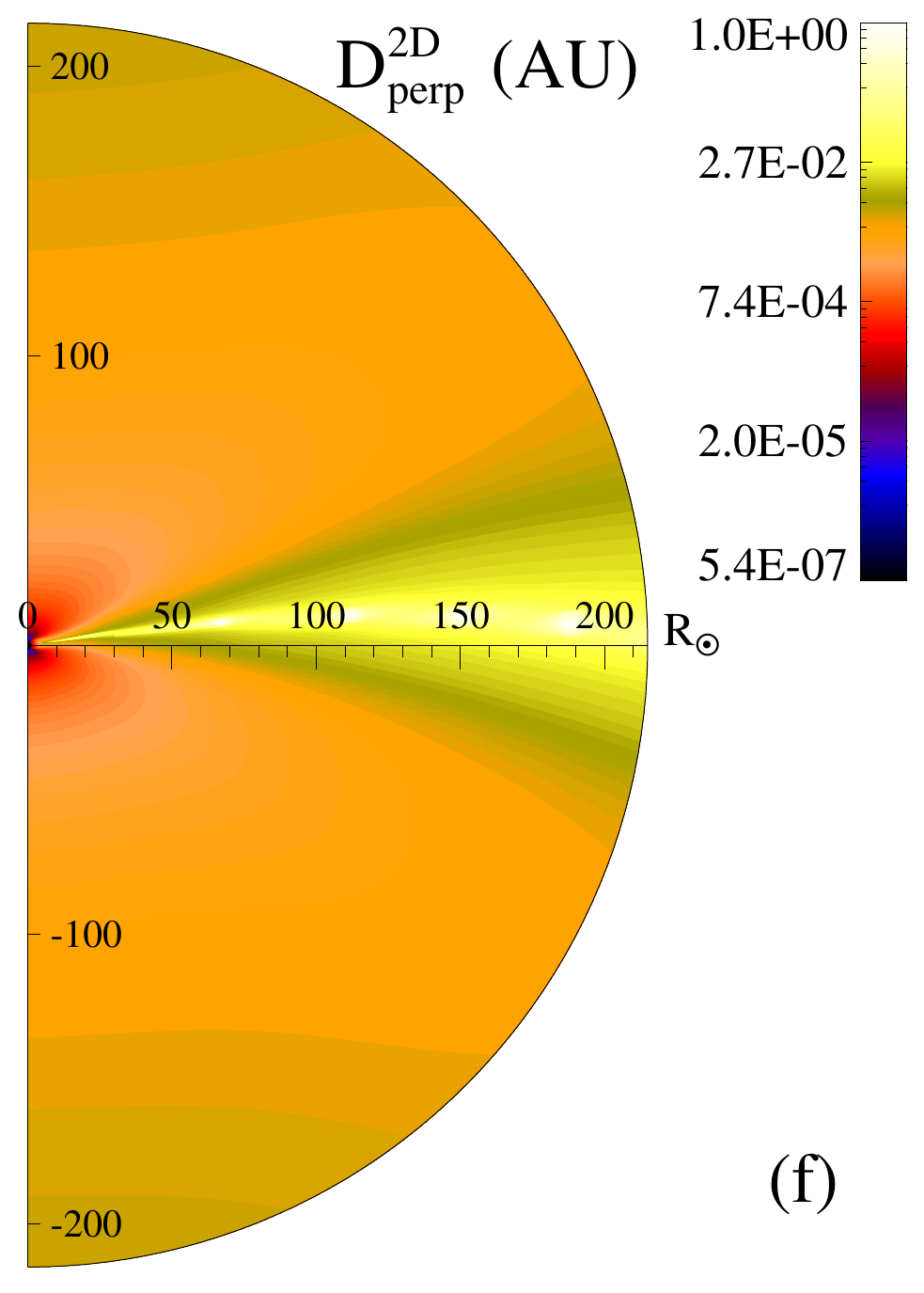}{.245\textwidth}{}
\fig{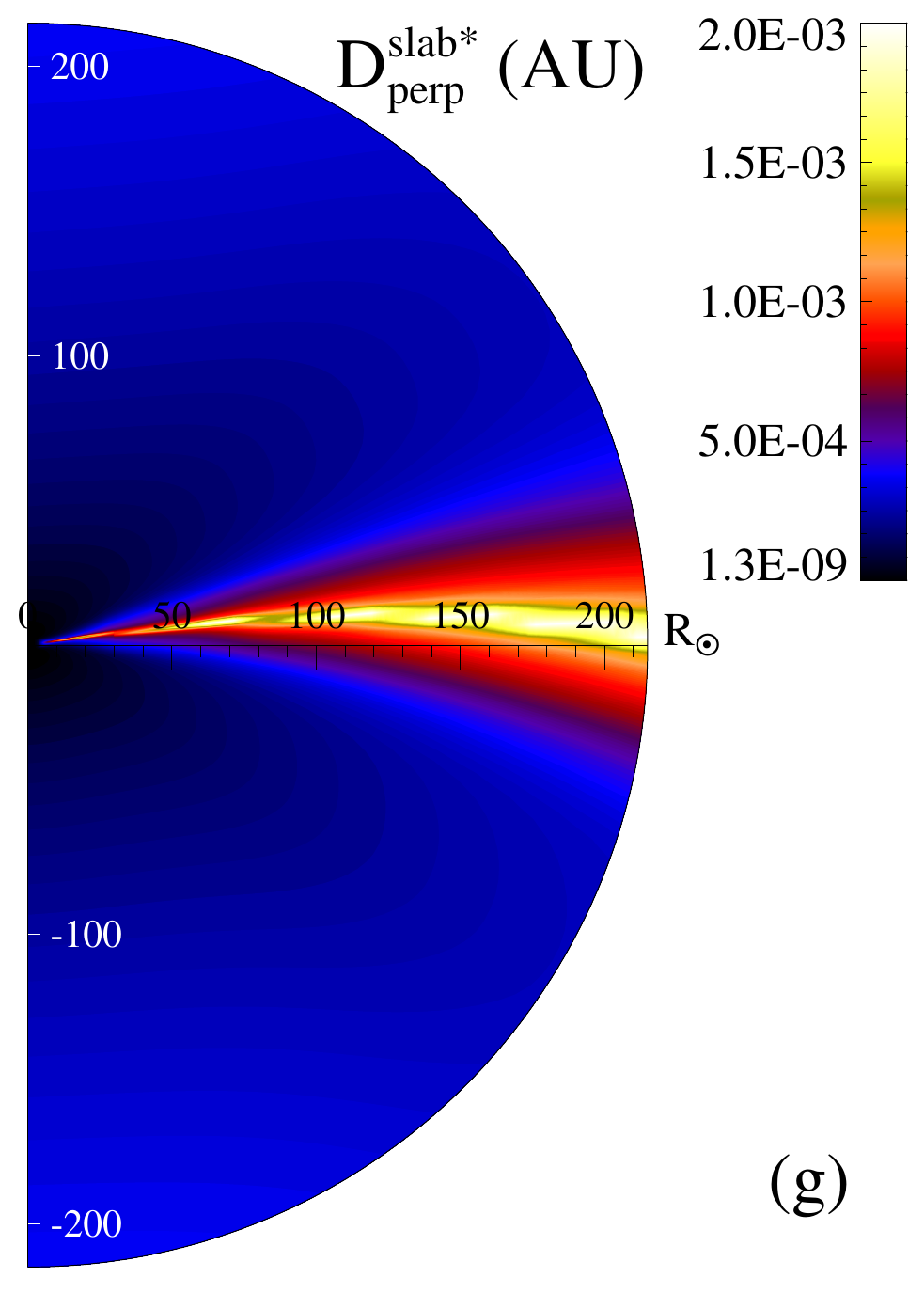}{.245\textwidth}{}
\fig{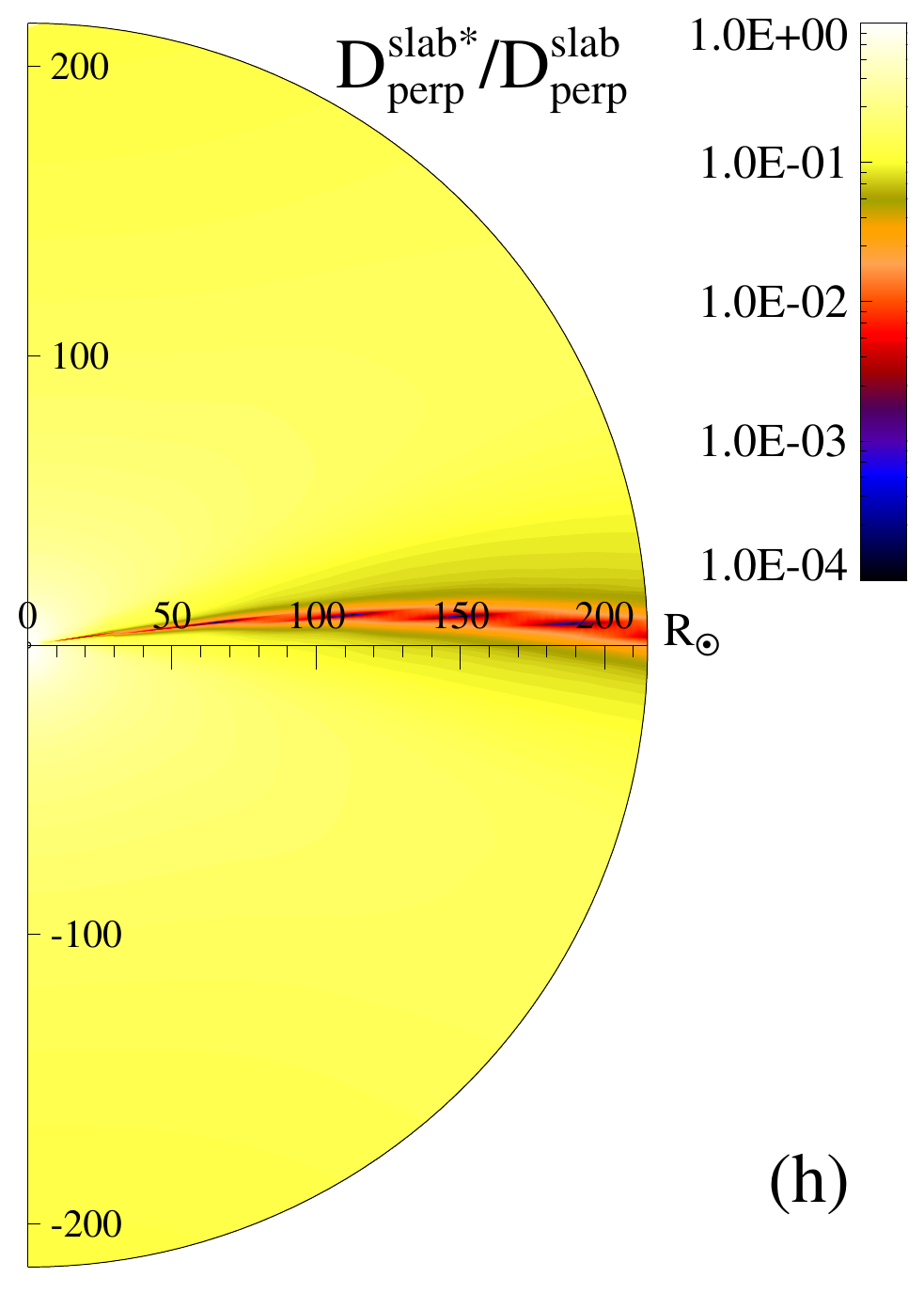}{.245\textwidth}{}
}
\caption{Panels (a)-(d) show global simulation results for the mean magnetic-field strength \(B\), proton number density \(n\), turbulence energy \(Z^2\), and the correlation scale \(\lambda\), respectively, in a meridional plane at 1\degree~heliolongitude and heliocentric distances of \(1\dash 215~\rs\), from Run I. These quantities are used to compute the field line diffusion coefficients \(\dslab\) and \(\dtwo\) using Equations \eqref{eq:Dslab} and \eqref{eq:D2d}, and the suppressed slab diffusion coefficient \(\dslabsup\) [Equation \eqref{eq:Dslab_sup}], shown in panels (e)-(g), respectively. Panel (h) shows the ratio \(\dslabsup/\dslab\).}
\label{fig:merid_plane}
\end{figure*}

The input parameters specified at the coronal base include: the driving amplitude of Alfv\'en waves (30 km~s$^{-1}$), the
density ($1 \times 10^8$ particles cm$^{-3}$), the correlation scale of turbulence (\(10,500\)~km), and temperature
($1.8 \times 10^6$~K). The cross helicity in the initial state is set as \(\sigma_c = -\sigma_{c0} B_r/B_r^\text{max}\), where \(\sigma_{c0}=0.8\), \(B_r\) is the radial magnetic field, and \(B_r^\text{max}\) is the maximum absolute value of \(B_r\) on the inner boundary. The magnetic field magnitude is assigned using a source magnetic dipole (with strength 12 G at the poles to match values observed by Ulysses). The input parameters also include the fraction of turbulent energy absorbed by protons $f_p = 0.6$. Further details on the numerical approach and initial and boundary conditions may be found in \cite{usmanov2018}.

\section{Results}\label{sec:results}

We begin this section by presenting a sample of the global simulation results that we use to evaluate field line diffusion: the large-scale magnetic field magnitude \(B\), proton density \(n\), turbulence energy \(Z^2\), and correlation scale \(\lambda\). Figures \ref{fig:merid_plane}(a)-(d) show sample results from Run I in a meridional plane at 1\degree~heliolongitude, in heliographic coordinates \citep[HGC;][]{franz2002pss}. The heliospheric current sheet (HCS) is clearly visible at low-latitudes in panel (a), and it is also a region of relatively high density, as expected for low-latitude slow wind during solar minimum \citep{mccomas2000JGR}. The turbulence quantities also exhibit interesting (and expected) latitudinal variation. Mid-to-high latitude fast wind is more Alfv\'enic, with higher turbulence energy relative to lower latitudes at the same radial distance. As expected the turbulence energy density decreases
with radius almost everywhere. 
On the other hand, the correlation scale increases with distance from the Sun as the turbulence ``ages'' and the inertial-range cascade extends to larger spatial scales \citep{matthaeus1998JGR,bruno2013LRSP,deforest2016ApJ828,chhiber2018apjl}, with the more turbulent slow-wind ageing faster. For more detailed description of the base variables from the simulation see \cite{usmanov2018}.

Next, to evaluate the field line diffusion
coefficients according to Equations \eqref{eq:Dslab}, \eqref{eq:D2d}, and \eqref{eq:Dslab_sup}, we make some approximations that are appropriate to the inner heliosphere, for estimating the slab and 2D fractions of turbulent energy and correlation scales. First, we note
that the turbulent fluctuations in our model are primarily transverse to the mean magnetic field \citep{breech2008turbulence,usmanov2014three,usmanov2018}, while in addition the 
quasi-2D contribution to the energy is generally  the dominant contribution. Therefore we associate the correlation scale of the 2D turbulence with the dynamically evolving 
similarity scale $\lambda$ 
in our turbulence model, so that \(\lambda_\text{c2}=\lambda\). Observational studies indicate that the slab correlation scale is about a factor of two larger than the 2D correlation scale \citep{osman2007ApJ654,Weygand2009JGRA,weygand2011JGR116}, and accordingly, we assume \(\lambda_\text{slab} = 2 \lambda_\text{c2}\). Next, we assume that the slab fraction of the total magnetic fluctuation energy \(\langle B'^2 \rangle\) is \(f_s = 0.2\), a decomposition that is well supported both theoretically \citep{zank1992waves,zank1993nearly,Zank2017ApJ835} and observationally \citep{matthaeus1990JGR,bieber1994proton,
bieber1996dominant}. Then we have \(b^2_\text{s} = f_\text{s} \langle B'^2 \rangle\) and \(b^2_\text{2D} = (1-f_\text{s}) \langle B'^2 \rangle\). Finally, we convert \(Z^2\) from the simulation to \(\langle B'^2\rangle\) using the relation \(\langle B'^2\rangle =  4\pi\rho Z^2/(r_\text{A}+1)\) (see Section \ref{sec:model_desc}), and an Alfv\'en ratio \(r_\text{A} = \langle {v'}^2\rangle/\langle {b'}^2\rangle =1/2\) is assumed \citep{tu1995SSRv}.\footnote{\footnotesize{Refinements have been developed in recent years to this simplified perspective on the decomposition of slab and 2D fluctuation energies and the constant Alfv\'en ratio \citep[e.g.,][]{Hunana2010ApJ718,Oughton2011JGRA116,Zank2017ApJ835}. These studies find that the evolution of the two components is markedly different in the outer heliosphere (beyond \(\sim 3\) au) where driving by pickup ions leads to an increase in the slab component’s energy, while the energy of the 2D component continues to decrease with heliocentric distance. However, for the purposes of our present work focusing on the inner heliosphere, our simple decomposition into slab and 2D components using a constant ratio seems appropriate \citep[see discussion in Section 3 of][]{chhiber2017ApJS230}.}}

Using the above approximations, we are able to compute all parameters required to evaluate the diffusion coefficients, based on the dynamically evolved
turbulence parameters provided by the subgrid turbulence model in the {\it Usmanov et al.}
global MHD code. 
2D maps of the field line diffusion coefficients in meridional planes at 1\degree~heliolongitude are shown in panels (e)-(g) of Figure \ref{fig:merid_plane}. We see that \(\dperp\) increases with distance from the Sun, and is relatively larger in the slow-wind region around the HCS. The suppressed slab diffusion coefficient \(\dslabsup\) behaves similarly. It is interesting to examine the ratio \(\dslabsup/\dslab\) in panel (g), which reveals that there is significant suppression of slab diffusion for trapped field lines at all latitudes, with the strongest effect seen near the HCS.

With all the quantities needed to solve Equations \eqref{eq:ODE1} and \eqref{eq:ODE2} in place, we next trace magnetic field lines from the coronal base out to 1 au, employing   trilinear interpolation 
of the relevant quantities (\(B \equiv B_0,\dperp\), and \(\dslabsup\)) to the field line under consideration. The ODEs are then integrated along the large-scale field line using a second-order Runge-Kutta method to evaluate the mean-squared spread of field lines \(\langle \Delta R^2 \rangle\), with an initial spread corresponding to a circle of radius 2\degree~at 0.1 au. To estimate the filamentation distance and the associated ``suppressed'' spreading, we integrate Equation \eqref{eq:ODE2} with different values of initial \(\langle \Delta R_\ast^2 \rangle\), which represent varying initial rms displacements of field lines from the center of an O-point. We also vary the slab fraction \(f_s\) of the magnetic fluctuation energy. Stronger trapping is expected for smaller values of initial displacement and \(f_s\) (see Section \ref{sec:filam}).
%
%
%
%
%
A three-point Lagrangian interpolation formula \citep[e.g.,][]{hildebrand1974numerical} is used to approximate the derivative of \(B_0\) along \(z\). In the following two subsections we describe the results obtained in this way 
from Runs I and II. In the following, longitude (latitude) is used interchangeably with heliolongitude (heliolatitude), referring to HGC coordinates.

%
\subsection{Run I: Dipole-based Simulation}\label{sec:dipole}
We recall 
that Run I is a dipole-based run (\(1~R_\Sun\dash 5\) au) with the dipole axis tilted by 10\degree~relative to the solar rotation axis. Integration of Equations \eqref{eq:ODE1} and \eqref{eq:ODE2}, with a starting spread corresponding to a circle of given radius at 0.1 au provides values for the rms transverse displacement
\(\langle \Delta R^2 \rangle^{1/2}\)  and rms suppressed slab displacement \(\langle \Delta R_\ast^2 \rangle^{1/2}\), 
as described in Sections \ref{sec:flrw} and \ref{sec:filam}.
The latter quantity is 
relevant to 
the spread of field lines initially trapped in small-scale topological structures where slab diffusion is suppressed. 

\begin{figure}
\includegraphics[width=.5\textwidth]{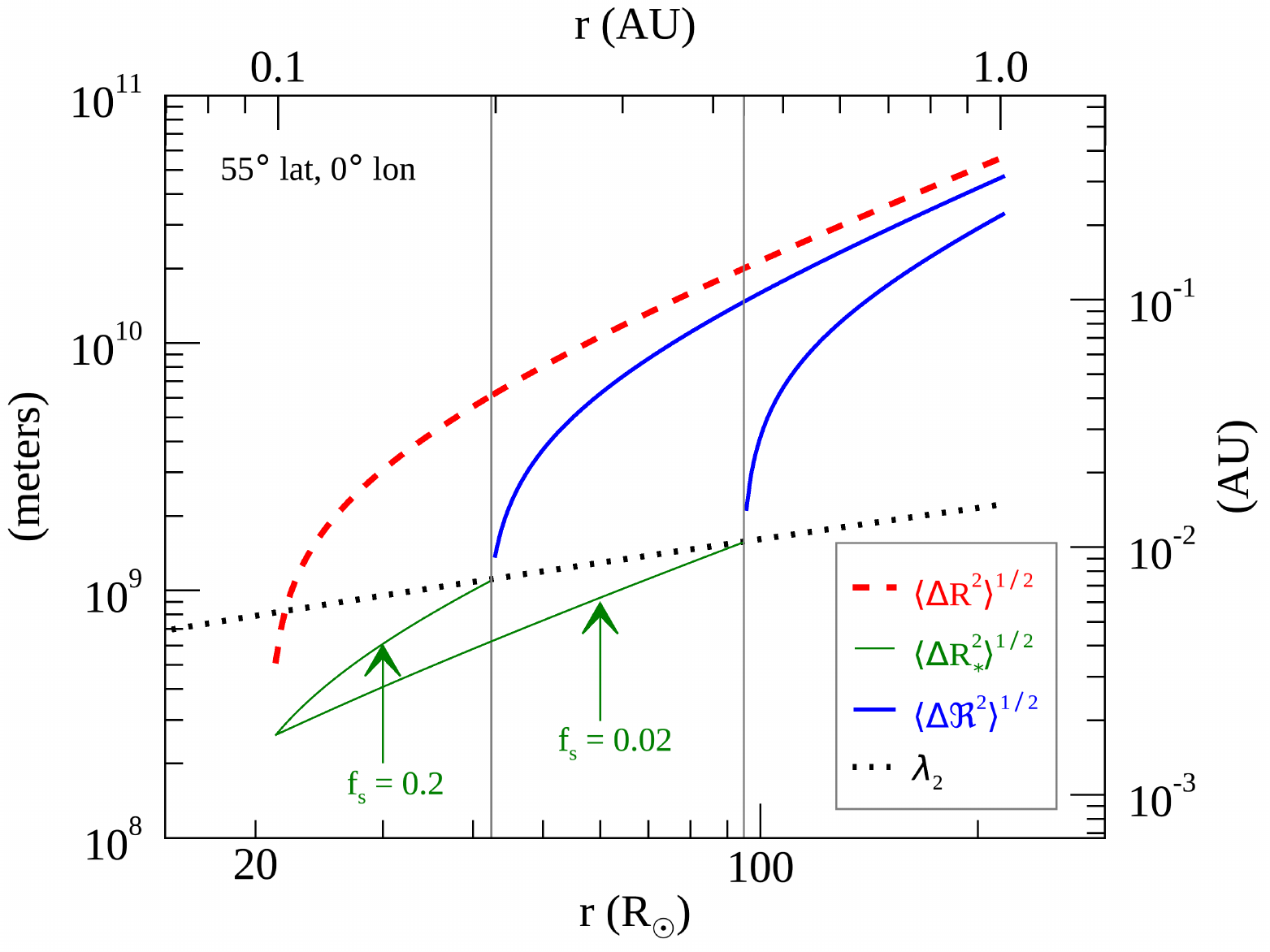}
\caption{Results from Run I, a global simulation for a solar magnetic dipole with  10\degree~tilt, showing the 2D bendover scale \(\lambda_2\) (dotted black curve), normal FLRW spread \(\langle \Delta R^2 \rangle^{1/2}\) (dashed red curve), suppressed spread \(\langle \Delta R_\ast^2 \rangle^{1/2}\) (solid green curves), and spread after escape from trapping boundary \(\langle \Delta \mathscr{R}^2 \rangle^{1/2}\) (solid blue curves), along a field line that originates at 55\degree~latitude and 0\degree~longitude. The latter two quantities are shown for two values of slab energy fraction -- $f_s=0.2$ and $f_s=0.02$, with initial \(\langle \Delta R_\ast^2 \rangle^{1/2}\) corresponding to \(2.7\times 10^8\) m. Vertical grey lines indicate the filamentation distance for the two cases, at \(42~\rs\) and \(95~\rs\), respectively. For the calculation of normal FLRW spread \(\langle \Delta R^2 \rangle^{1/2}\), \(f_s=0.2\).}
\label{fig:delx_2}
\end{figure}

Figure \ref{fig:delx_2}(a) shows the
above two quantities for a field line that originates at 55\degree~latitude and 0\degree~longitude. Here an initial spread corresponding to a circle of radius 2\degree\ is used for evaluating \(\langle \Delta R^2 \rangle^{1/2}\), while an intial spread of 1\degree\ is used for evaluating \(\langle \Delta R_\ast^2 \rangle^{1/2}\). Also shown is the 2D bendover scale \(\lambda_2\).
The filamentation distance \(r_f\) (vertical grey line) is the heliocentric distance at which \(\langle \Delta R_\ast^2 \rangle^{1/2} = \lambda_2\), the 2D bendover scale.
Beyond this point,
field lines that are initially trapped can begin to undergo an unsuppressed random walk. We also show \(\langle \Delta \mathscr{R}^2 \rangle^{1/2}\), which represents the spreading of field lines after they have escaped trapping boundaries (see schematic in Figure \ref{fig:cartoon}). This quantity is obtained by integrating the normal FLRW equation [Equation \eqref{eq:ODE1}], starting at the filamentation distance 
\(r_f\) with an initial spread equal to the value of \(\langle \Delta R_\ast^2 \rangle^{1/2}\) at \(r_f\). Note that we only show mid-latitude field lines from Run I since all low-latitude field lines close back onto the solar surface for the dipolar-source magnetic field; however, even such mid-latitude field lines can meander to the ecliptic plane, as 
 discussed below. Field lines originating at low latitude may be obtained from the magnetogram-based Run II (see Section \ref{sec:magnetogram}, below).


The results in Figure \ref{fig:delx_2} show an rms spread of around 3-6 \(\times 10^{10}\) m at 1 au for both 
regular, untrapped field lines and for those that are initially trapped. The latter experience significantly reduced diffusion up to the filamentation distance (\(42~\rs\) and \(95~\rs\)\ for the two slab fractions shown), and then begin to approach the outer envelope. The corresponding angular spreads near Earth are around 25\degree. Note that the angular spread refers to the angle subtended 
by the radius of rms diffusive spreading at a distance of 1 au. Therefore the separation in longitude between two field lines that have random walked in opposite directions can be \(\sim 50\degree\)\ (see also Section \ref{sec:magnetogram}). These spreads are consistent with those inferred from observations of SEPs from some impulsive solar events \citep[e.g.,][]{wibberenz2006ApJ,wiedenbeck2013ApJ,droge2014JGR} and from full particle orbit simulations by \cite{tooprakai2016ApJ} (see also Figure \ref{fig:paisan}).
It is interesting to note that \citet{laitinen2016AA}
find a similar angular spread of SEPs in the distance range of 0.5 to 1.5 au, in a calculation in which it is shown that the FLRW makes a substantial contribution, although in that case the field line spread is entirely of a diffusive nature. 

\begin{figure}
\includegraphics[scale=.5]{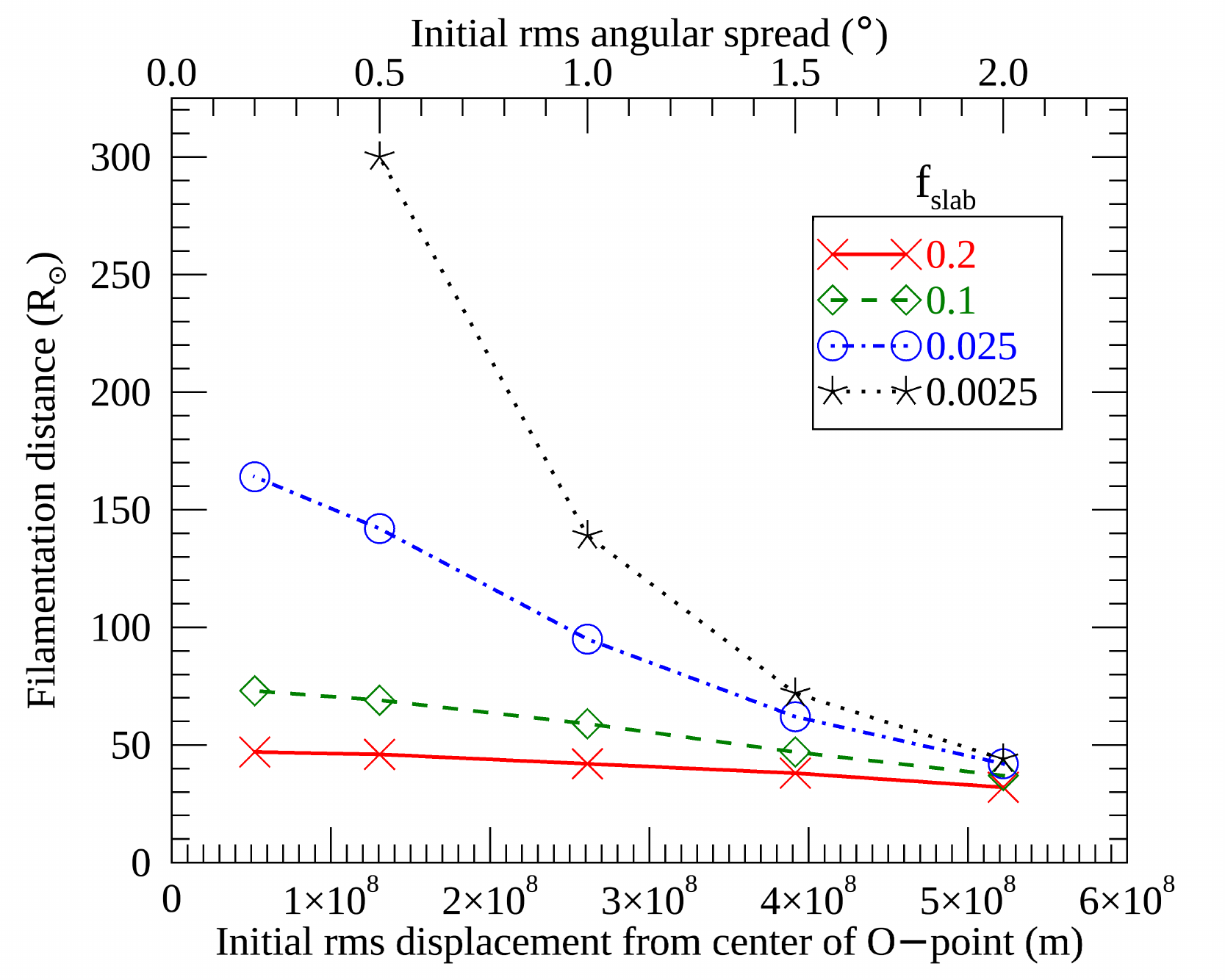}
\caption{Parameter study for calculation of filamentation radius $r_f$ for a field line from the  tilted-dipole simulation (Run I), at initial latitude and longitude of 55\degree\ and 0\degree, respectively. The figure shows \(r_f\) for different values of slab-energy fraction \(f_s\) and varying initial rms displacement from the center of the O-point [i.e., initial \(\langle \Delta R_\ast^2 \rangle^{1/2}\) in Equation \eqref{eq:ODE2}], for the case of field lines initially trapped in a magnetic island.}
\label{fig:r_filam}
\end{figure}

As we have calculated it, the filamentation distance is explicitly dependent on the amplitude of slab fluctuations present in the turbulence. In addition, we recall that \cite{chuychai2007ApJ} found that the suppressed diffusive-escape theory\footnote{\footnotesize{Used here to compute the suppressed slab diffusion coefficient; see Section \ref{sec:filam}.}} matches simulations best for small slab fraction (0.001-0.0025), i.e., when the 2D component is dominant \citep[see Figures 8 and 9 of][]{chuychai2007ApJ}. It is therefore of interest to examine, within the context of that theory, how the filamentation distance varies in the present framework, as one varies the slab fraction and the distance of a field line from the central O-point in a flux tube. A brief parameter study of these effects is illustrated in Figure \ref{fig:r_filam}. Here the initial displacement from the O-point is varied from 5 $\times 10^7$~m to 5.2 $\times 10^8$~m, and the slab fraction is varied from 0.0025 to 0.2 of the total magnetic fluctuation energy. For each value of the slab fraction the filamentation distance monotonically increases as the field line's initial position is migrated closer to the O-point \citep[see also Figure 5 of ][]{chuychai2007ApJ}. In addition, there is a dramatic increase in filamentation distance for smaller slab fractions. 

\begin{figure}
\centering
\includegraphics[width=.5\textwidth]{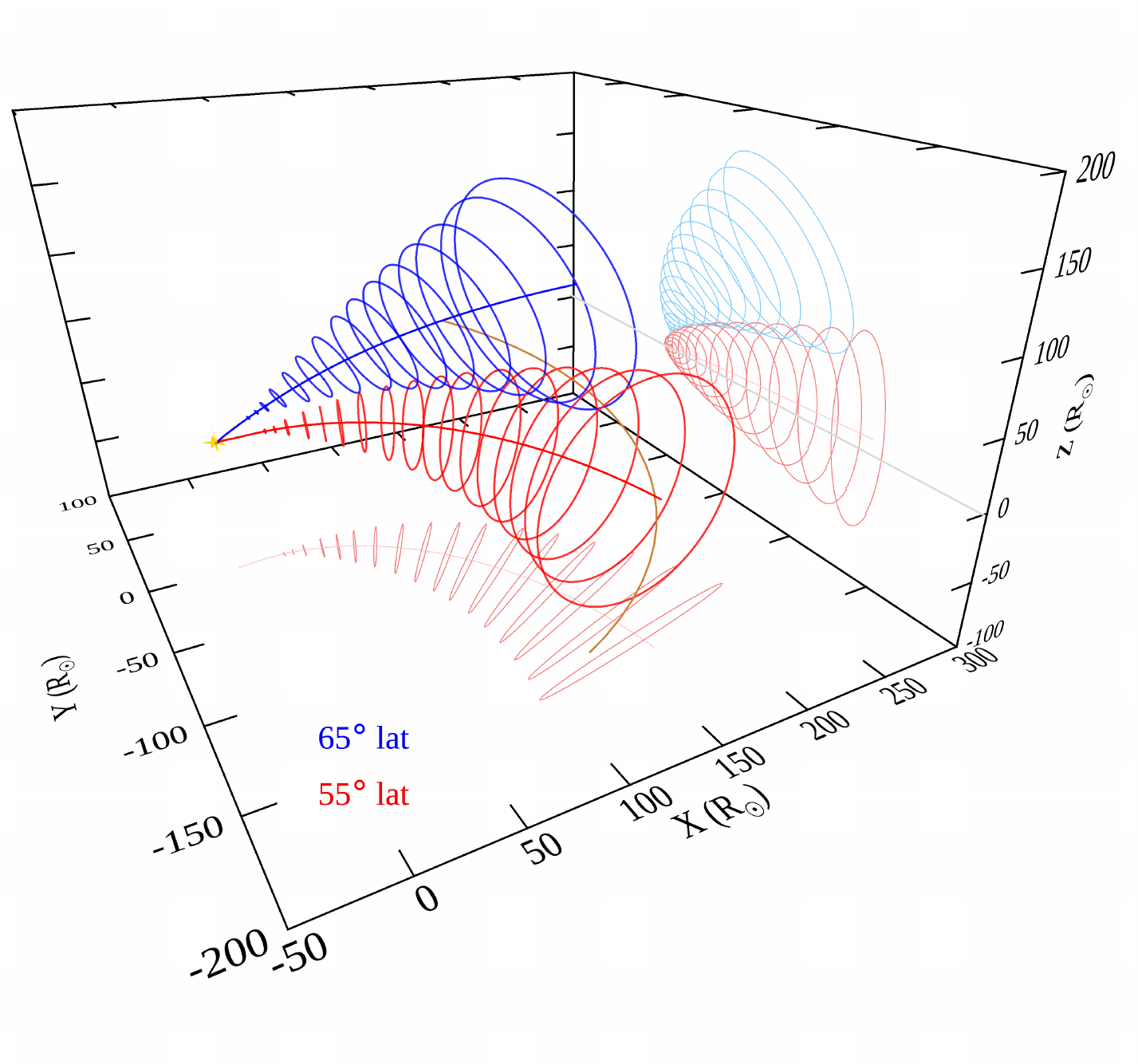}
\caption{3D representation of two field lines originating at 65\degree\ and 55\degree\ latitude, and 0\degree\ longitude, traced from the coronal base at \(1~\rs\) to the distance of Earth orbit (depicted nominally as a brown curve) for global simulation Run I. 
The circles around the central field lines have a radius equal to \(\langle \Delta R_\ast^2 \rangle^{1/2}\) from 0.1 au up to the filamentation distance and a radius \(\langle \Delta \mathscr{R}^2 \rangle^{1/2}\) beyond that. The projections of the 55\degree~case on the X-Y and Y-Z planes, and a  projection of the 65\degree\ case on the Y-Z plane, are shown in muted colors. The light-grey line in the Y-Z plane marks Z=0 for reference. The axes represent HGC coordinates.}
\label{fig:3d_2}
\end{figure}

\begin{figure}
\gridline{\fig{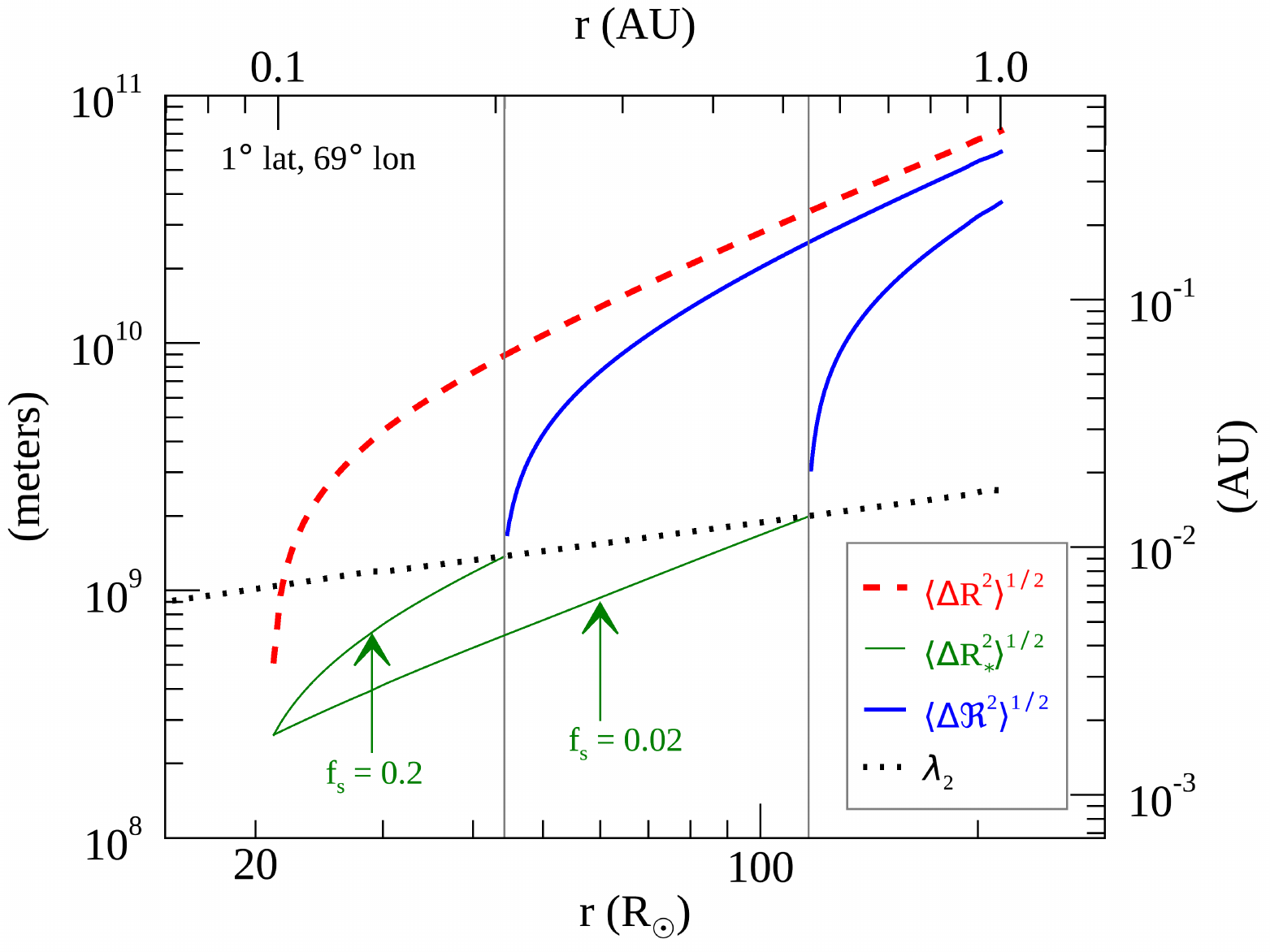}{.45\textwidth}{(a)}}
\gridline{\fig{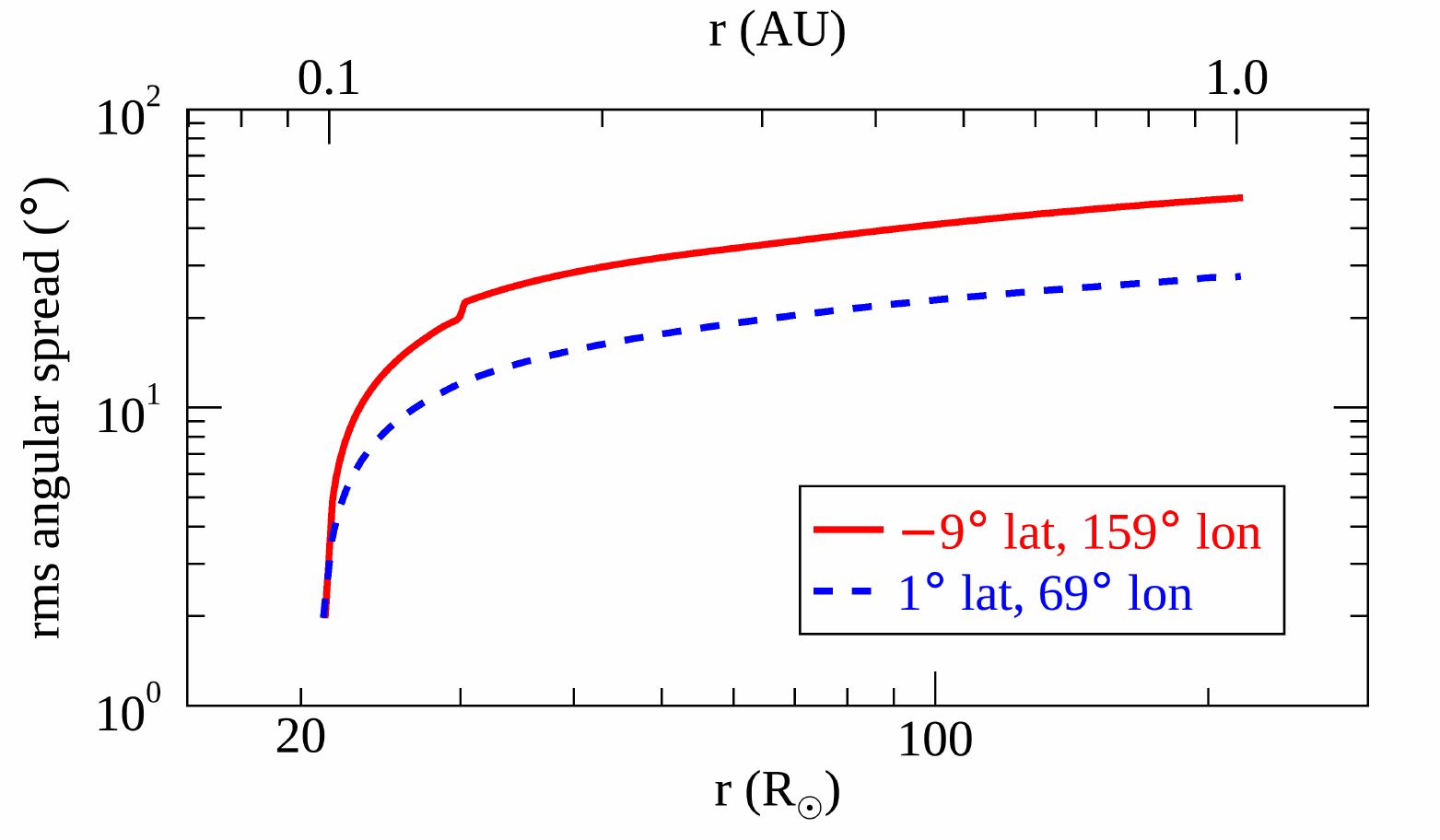}{.45\textwidth}{(b)}}
\caption{Results from Run II, a global simulation based on a CR 2123 magnetogram.
(a) 2D bendover scale \(\lambda_2\) (dotted black curve), normal FLRW spread \(\langle \Delta R^2 \rangle^{1/2}\) (dashed red curve), suppressed spread \(\langle \Delta R_\ast^2 \rangle^{1/2}\) (solid green curve), and spread after the end of suppression \(\langle \Delta \mathscr{R}^2 \rangle^{1/2}\) (solid blue curves), along a field line that originates at 1\degree~latitude and 69\degree~longitude. The latter two quantities are shown for two values of slab energy fraction -- $f_s=0.2$ and $f_s=0.02$, with initial \(\langle \Delta R_\ast^2 \rangle^{1/2}\) corresponding to \(2.7\times 10^8\) m. Vertical grey lines indicate the filamentation distance for the two cases, at \(44~\rs\) and \(120~\rs\), respectively. For the calculation of normal FLRW spread \(\langle \Delta R^2 \rangle^{1/2}\), \(f_s=0.2\). (b) Angular spread of two selected field lines as a function of heliocentric distance. The 1\degree~and -9\degree~latitude fieldlines originate at 69\degree~and 159\degree\ longitude, respectively. The rms angular spread is computed as \(\theta = \langle \Delta R^2 \rangle^{1/2}/r\), where \(r\) is the radial coordinate of a field line.}
\label{fig:delx_3}
\end{figure}
%

%
%

%
\begin{figure}
\includegraphics[width=.5\textwidth]{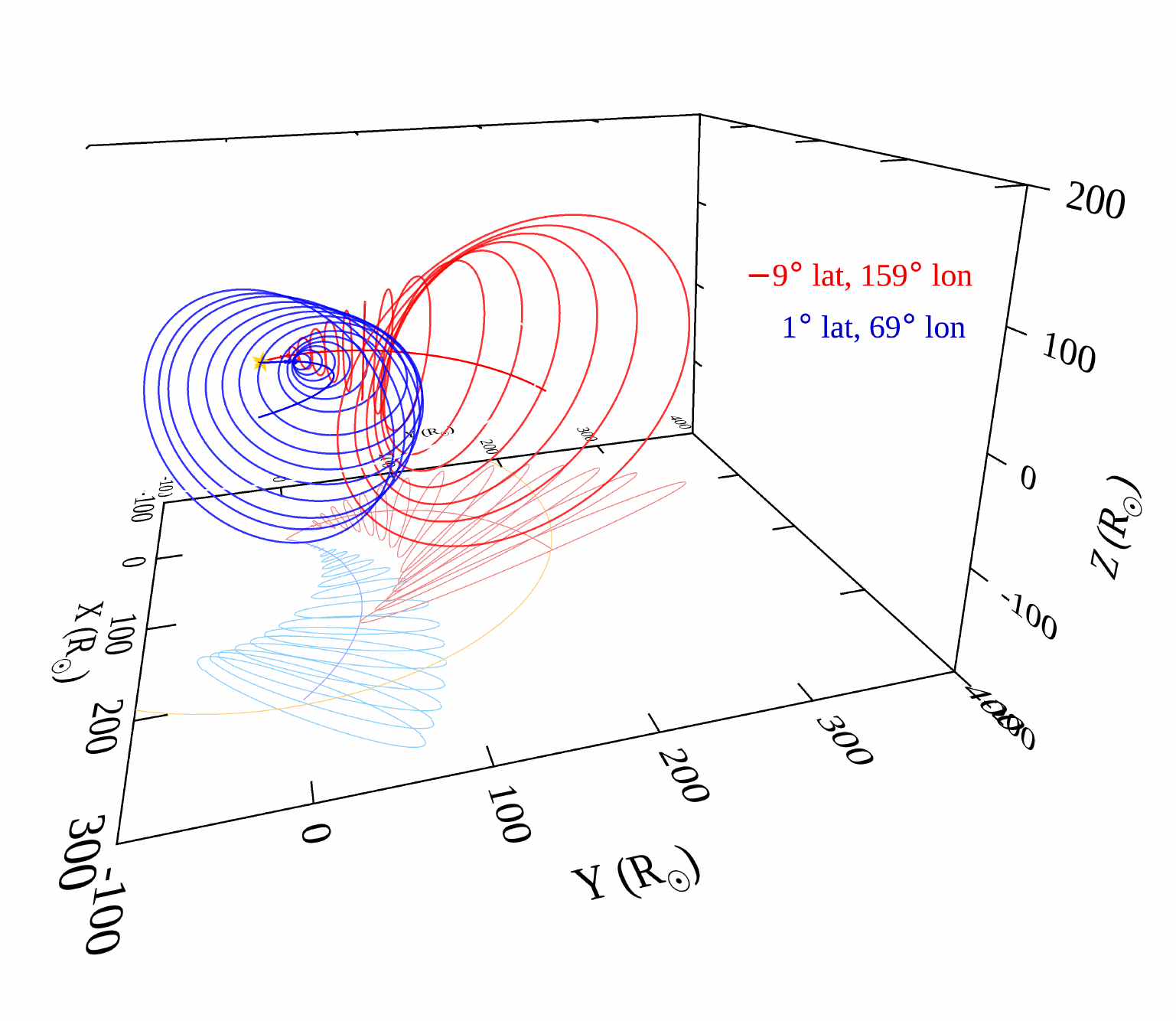}
\caption{3D representation of two field lines originating at latitude/longitude of 1\degree/69\degree\ (blue) and -9\degree/159\degree\ (red), traced from the coronal base at \(1\rs\)\ to the distance of a nominal Earth orbit (depicted as a pale brown curve in the X-Y plane projection), from global simulation Run II. The circles around the central field lines have a radius equal to \(\langle \Delta R_\ast^2 \rangle^{1/2}\) from 0.1 au up to the filamentation distance, and a radius \(\langle \Delta \mathscr{R}^2 \rangle^{1/2}\) beyond that.
The projections of the field lines on the X-Y plane are shown in muted colors. The axes represent HGC coordinates.}
\label{fig:3d_4}
\end{figure}

Figure \ref{fig:3d_2} shows a 3D representation of two field lines, from the solar surface to Earth orbit (depicted as a brown curve). The circles around the central field lines have a radius equal to \(\langle \Delta R_\ast^2 \rangle^{1/2}\) from 0.1 au up to the filamentation distance, and a radius \(\langle \Delta \mathscr{R}^2 \rangle^{1/2}\) above it. At 1 au, the rms separation of random-walking field lines can be in excess of \(100~\rs\). Further, one observes from the Y-Z projection of the field line originating at 55\degree\ latitude, that field lines originating from solar mid-latitudes can meander to the ecliptic by their arrival at 1 au. The projection of the two field lines on the Y-Z plane shows that there can be mixing between field lines originating at different latitudes. The next subsection shows a similar figure for the case of the magnetogram-based simulation.


%
\subsection{Run II: Magnetogram-based Simulation}\label{sec:magnetogram}
In this subsection we present results from Run II, which employs a magnetogram corresponding to CR 2123 (2012 April 28 -- 2012 May 25), 
a period of high solar activity. 
Due to the more complex and more realistic magnetic field produced by the magnetogram initialization,
it is of interest to consider two problems: 
First we will integrate outward to 1 au
from a specified source region, as in the previous section. 
Next, we will begin an integration 
at 1 au and integrate inward along a large-scale field line to infer a broader area that represents a possible source region. 

To begin, as in the dipolar case, we 
integrate 
Equations \eqref{eq:ODE1} and \eqref{eq:ODE2} starting
at 0.1 au, with a specified 
initial  spread corresponding to a circle of radius 2\degree\ (1\degree\ for the case of suppressed spread). Results based on this run are shown in Figures \ref{fig:delx_3} -- \ref{fig:3d_4}. Note that in this case
we are able to examine field lines originating at low latitudes, 
in contrast to the dipole run wherein all low-latitude field lines close back onto the solar surface. 

The results are qualitatively similar to those in the previous subsection, but the rms spreads are almost a factor of two larger, possibly owing to stronger fluctuations during solar maximum.
It is noteworthy 
that in this magnetogram-driven case, the typical angular spread of field lines at 1 au 
from the source can exceed 50\degree\ [see Figure \ref{fig:delx_3}(b)].
Likewise, examining
Figure \ref{fig:3d_4}, it is possible to observe an 
rms separation of field lines at 1 au that is in excess of 200 Earth radii. Figure \ref{fig:3d_4} also shows that mixing can occur between field lines originating at widely separated (\(\sim 90\degree\)) longitudes, due to FLRW.

\begin{figure}
\includegraphics[scale=.5]{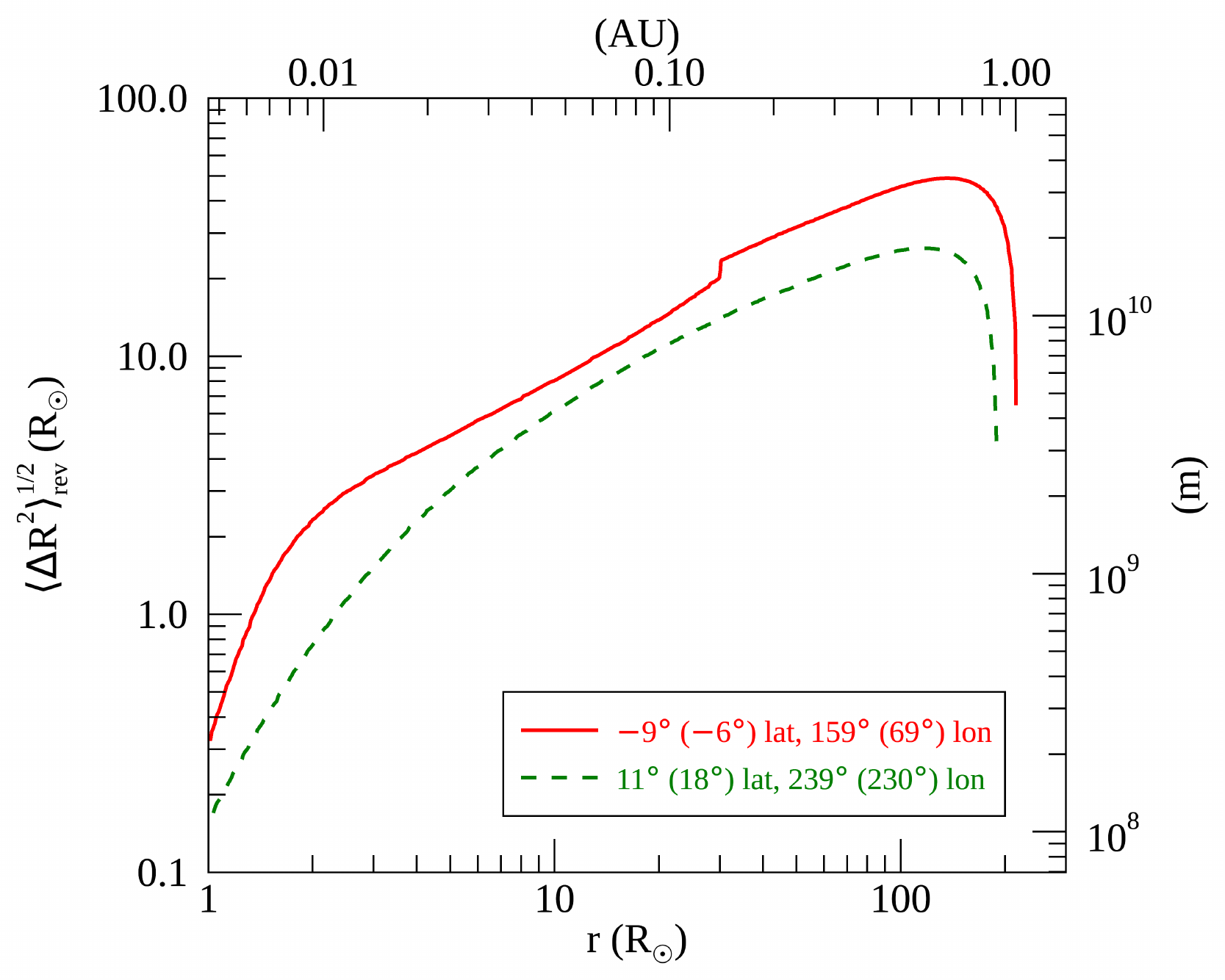}
\caption{The reverse FLRW spread \(\langle \Delta R^2\rangle^{1/2}_\text{rev}\) of selected field lines from global simulation Run II, assuming an initial spread of zero at 1 au and integrating the FLRW equation \eqref{eq:ODE1} Sunward. 
The solid red and dashed green curves show the rms spread around field lines that originate at latitude/longitude of \(-9\degree/159\degree\)~ and 11\degree/239\degree\ near the solar surface, and at 1 au have a latitude/longitude of \(-6\degree/69\degree\)~ and \(18\degree/230\degree\), respectively.}
\label{fig:reverse1}
\end{figure}
\begin{figure*}
\centering
\includegraphics[scale=.18]{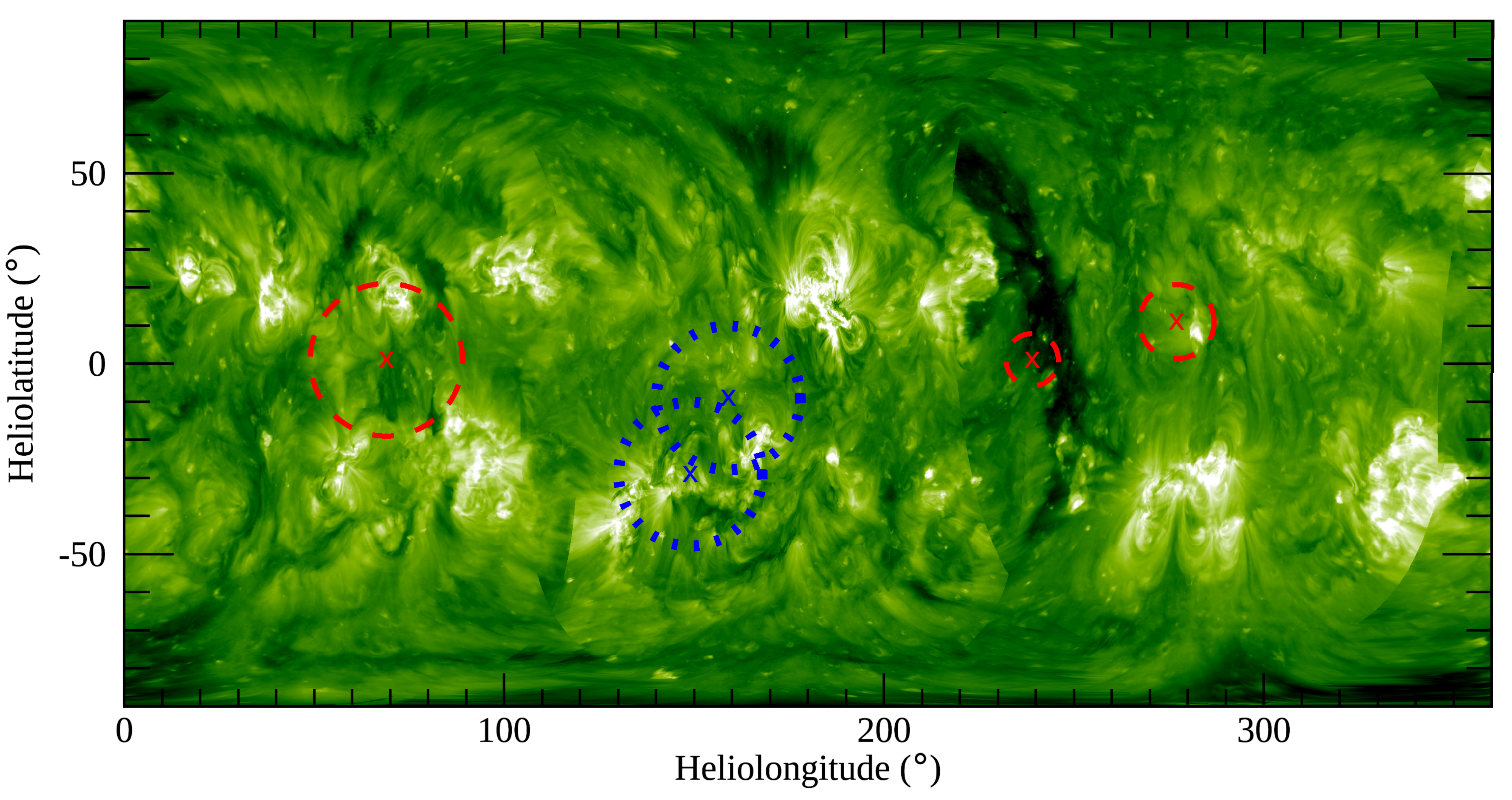}
\caption{Synoptic map of the solar corona dated 2012 May 12, during CR 2123. The synchronized EUV image is based on \textit{STEREO}/EUVI A and B 195 \AA\ and \textit{SDO}/AIA 193 \AA\ observations. The processing involved in producing this image is described in \cite{caplan2016ApJ}. From our global simulation Run II based on a CR 2123 magnetogram, we calculate the reverse spread of field lines connected to specific locations at 1 au from the Sun, as described in the text. The blue and red `$\times$' symbols mark the origins of selected field lines in low-latitude regions with positive and negative polarities, respectively. The circles (dashed red for negative and dotted blue for positive polarities) have radii equal to \(\langle \Delta R^2\rangle^{1/2}_\text{rev}\) at the solar surface, and mark regions from which a field line, at the position at 1 au corresponding to that of the central `$\times$' field line, is likely to have originated. We show that because of turbulent fluctuations, there can be a significant uncertainty in back-tracing of field lines from 1 au. The synoptic map was obtained from \href{http://www.predsci.com/chd/}{http://www.predsci.com/chd/}.
}
\label{fig:reverse2}
\end{figure*}

Turning the problem around, and integrating inward, it is possible to investigate the extent of the spatial region on the solar surface from which a field line localized at 1 au could have originated. For this purpose we integrate the FLRW equation \eqref{eq:ODE1} Sunward, assuming an initial spread of zero at 1 au. In Figures \ref{fig:reverse1} and \ref{fig:reverse2} we show the results of this computation of the reverse FLRW for selected field lines. One observes that the region of likely field line spreading extends over regions comparable to the large features on the solar synoptic map, on the order of the supergranulation scale of 
$\sim$ 35,000 km, or even larger. 
Thus we see that 
turbulent fluctuations
can induce 
a significant uncertainty (around 20\degree\ in latitude/longitude) 
in any sort of back-tracing of field lines to the solar surface. 
Field line random walk
due to unresolved fluctuations
introduces an inherent 
inaccuracy 
in determination of the 
location on the Sun that is magnetically connected to a location of interest at distance $\sim$ 1 au. This translates directly into uncertainty
in 
identifying coronal sources 
of solar wind plasma and energetic particles.

\section{Discussion}\label{sec:disc}

The purpose of this paper has been to 
develop methods for approximating the 
spread of magnetic field lines in an 
inhomogeneous expanding medium such as the solar wind, and to implement that method 
employing data from a global heliospheric 
MHD model. The framework developed combines two contributors to magnetic field line spread. The first, a coherent effect, is due to expansion 
of the large-scale or resolved field, 
for which an explicit representation is available, e.g., from a global simulation.
The second effect included is the random 
influence on spreading due to unresolved turbulence that is known only in terms of its statistical properties such as its spectrum. 
The latter effect requires a theoretical 
model for how turbulence descriptors such as 
the variance and correlation scale 
vary along the resolved field line that is traced. Here we employ one particular approach for this step, namely implementing and solving a set of turbulence transport equations
for turbulence energy and correlation scale
that are self-consistently solved along 
with the large scale resolved 
``macro'' MHD model
\citep{usmanov2018}. 
Taken together, the approximate treatment of this problem that is developed in Section 
\ref{sec:formal} provides a fairly general 
framework for approximating the spread of magnetic field lines in a variety of problems. 
The main motivation here is the tracing of connectivity from solar sources to interplanetary space and vice versa. This has potential important implications for solar energetic particle propagation, as well as space weather applications. 

In particular,
embedding the above 
formalism into a global model 
that supplies both 
large scale magnetic fields and turbulence parameters
comprises 
an approach that 
shows promise in incorporating a degree of realism into 
quantifying field line connectivity and transport of SEPs. 
We are aware that some prior studies 
have 
evaluated the impact of 
field line meandering on particle transport,
and concluded that the impact is minimal. 
\citet{nolte1973SoPh} provide a concise summary of
several such studies, in each case 
finding no support for spreading due to FLRW in observations.
One such line of reasoning 
argues that field line meandering due to photospheric motions has sometimes been overestimated \citep{nolte1973SoPh,kavanagh1970RevGeoSp}.
In this regard we note that the time dependent boundary effect alluded to in this argument is actually distinct from our starting point, which is the standard static 
FLRW that emerges from spatially distributed fluctuations without relying on any specific boundary variations in time
\citep[however, see][for boundary models for generation of fluctuations]{giacalone2006ApJ}. More direct observational limits on transverse diffusion have also been reported based on 
multiple point observations of SEPs \citep{krimigis1971jgr}.
This conclusion is largely based on examples of observations in which 
intensity profiles remain relatively intact over significant distances. 
It is not possible to discount such reports; however, we have seen
that SEPs can indeed remain concentrated within 
trapping structures for relatively large distances, in some cases 
as far as 1 au. 
This phenomenon has been argued to be the basis for observed dropouts, which we attribute to topological trapping that delays the transition to standard diffusion. In the present study 
this effect is incorporated into the model by defining 
the filamentation distance, a quantity that provides an estimate of the typical distance over which this trapping occurs. 
Further quantitative observational study will be needed to 
ascertain the accuracy of the present theoretical framework.  

The results of the implementation of our framework within the global solar wind model are summarized below. We find rms spreads at 1 au of about \(10^{10}\) - \(10^{11}\) m, which correspond to rms angular spreads of about 20\degree\ - 60\degree; these estimates are generally consistent with the turbulence simulations of \cite{tooprakai2016ApJ}. The spreading is generally larger in the magnetogram-driven simulation corresponding to solar maximum, compared with the 10\degree\ dipole-tilt simulation. Note that this 
rms spreading implies that random-walking field lines 
originating within the same source region can develop separations as wide as 120\degree\ in longitude. Thus, our framework can account for the wide spreads that are sometimes seen in impulsive SEP observations \citep[e.g.,][]{wibberenz2006ApJ,wiedenbeck2013ApJ,droge2014JGR}. We also evaluate the heliocentric distances up to which field lines originating in  magnetic islands can remain strongly trapped; depending on the fraction of slab energy in the magnetic turbulence, and the initial distance from the center of the trapping island; this ``filamentation distance'' is estimated to be between 40 - \(150~\rs\), which is again consistent with \cite{tooprakai2016ApJ}. Finally, we estimate the uncertainty in magnetic connectivity of 1 au observations to solar sources; this is found to be around 20\degree\ in latitude/longitude at the source.

In closing, we emphasize 
that there is a reasonable degree of generality in the 
formalism developed here 
in Section \ref{sec:formal}, and that it may be employed 
within other global solar wind models and large-scale or global 
models of other astrophysical systems
for which the required 
turbulence parameters can be deduced.
To facilitate
applications, Appendix \ref{app:simple}
provides an implementation of 
our theoretical approach 
based on an extremely
simplified empirical treatment 
of the required parameters describing the magnetic field. Results of that exercise are illustrated in Figure \ref{fig:app_simple}.  

Although this approach may be adapted rather directly to different applications, we should recall that this theoretical development is approximate, given simplifications such as the neglect of correlations between the deflections due to large-scale expansion and the deflections due to the modeled small-scale turbulence. It may be of particular interest to extend this approach to the outer heliosphere and the very local interstellar medium, locations where field line connectivity may be relevant to explaining observations by \textit{Voyager}, \textit{IBEX}, and the upcoming \textit{IMAP} missions
\citep{Mccomas2018SSR,zank2019ApJ,zirnstein2020ApJ,fraternale2020ApJ}.  


\acknowledgments 
This research has been supported 
 in part by 
the  NASA LWS program  (NNX17AB79G) and HSR program (80NSSC18K1210 \& 80NSSC18K1648) and grant RTA6280002 from Thailand Science Research and Innovation,
and by
the Parker Solar Probe mission 
and the IS\(\odot\)IS~project 
 (contract NNN06AA01C) and a subcontract 
 to University of Delaware from
 Princeton University (SUB0000165).  The synoptic map used in Figure \ref{fig:reverse2} was obtained from \href{http://www.predsci.com/chd/}{http://www.predsci.com/chd/}.


%
\appendix

\numberwithin{equation}{section}

\section{Relating the 2D Correlation Scale to the 2D Bendover Scale}\label{sec:app}
%
To obtain a relationship between the 2D bendover scale \(\lambda_2\) and the 2D correlation scale \(\lambda_\text{c2}\) (see Section \ref{sec:filam}), we begin with the following form for the 2D modal power spectrum \citep{matthaeus2007spectral}: 
\begin{equation}
S_2(k_\perp) =  \left\{
 \begin{array}[c]{l}
    C_2\langle b_\text{2D}^2\rangle \lambda_2^2   
    (\lambda_2 k_\perp)^p, \\
    C_2\langle b_\text{2D}^2\rangle \lambda_2^2 (\lambda_2 k_\perp)^{-\nu -1},
 \end{array}
     \right.
 \begin{array}[c]{l}
    \text{if   } k_\perp \le 1/\lambda_2, \\
      \text{if   } k_\perp > 1/\lambda_2,
 \end{array}        \label{eq:2d_spec}
\end{equation} 
where \(C_2\) is a dimensionless constant, \(k_\perp\) is a 2D wavenumber, and \(p\) is a power-law index that determines the long-wavelength behavior of the spectrum. We also assume that the inertial range power-law index is \(\nu=5/3\), leading to an   omnidirectional spectrum ${\cal E}(k_\perp)=2\pi k_\perp S_2(k_\perp)\propto k_\perp^{-5/3}$ that is consistent with Kolmogorov theory. 
In the present work we assume \(p=2\), the lowest value that is consistent with strict homogeneity. We remark here that \cite{chhiber2017ApJS230} found that the value of \(p\) did not significantly influence energetic particle diffusion coefficients; however, if one uses a form for the spectrum with three separate wavenumber ranges \citep{matthaeus2007spectral}, then the diffusion coefficients may be sensitive to the choice for \(p\) \citep[see][]{engelbrecht2019apj}. 

The constant \(C_2\) may be determined by the normalization condition \(\langle b_\text{2D}^2\rangle = 2\pi \int_0^\infty S_2(k_\perp) k_\perp dk_\perp\), yielding \(C_2 = (\nu-1)(p+2)/[2\pi(p+\nu+1)]\). Next, we use the expression for the 2D correlation scale \citep[Equation (10) of][]{matthaeus2007spectral}:
\begin{equation}
\lambda_\text{c2} = \frac{\int [S_2(k_\perp)/k_\perp] dk_x dk_y}{\langle b^2_\text{2D}\rangle} = 
2\pi \frac{\int_0^\infty S_2(k_\perp) dk_\perp}{\langle b^2_\text{2D}\rangle}.
\end{equation}
On substituting the spectrum from Equation \eqref{eq:2d_spec} into the above and carrying out the integration, we obtain
\begin{equation}
\lambda_\text{c2} = \lambda_2 \frac{(p+2)(\nu-1)}{(p+1)\nu},
\end{equation}  
which yields \(\lambda_\text{c2} = 0.53 \lambda_2\) on choosing \(p=2\) and \(\nu=5/3\).


\section{Estimates of FLRW using simplified turbulence scalings}\label{app:simple}

To expand the applicability of our formalism to solar wind models that do not include a turbulence transport component, in this appendix we use simple approximations for turbulence parameters to obtain the FLRW diffusion coefficient. These approximations are (1) constant \(b/B_0\), and (2) the widely used \cite{Hollweg1986JGR} prescription for the turbulence correlation scale: \(\lambda \propto 1/B_0^{1/2}\). Here \(b\) and \(B_0\) are the rms magnetic fluctuation strength and the mean magnetic field, respectively. The Hollweg correlation scale is identified with the 2D turbulence. In Figure 12 the diffusion coefficient based on these approximations (and the corresponding rms spread of field lines) is compared with calculations based on the turbulence transport model. All calculations use \(B_0\) from the global solar wind simulation described in Section \ref{sec:sw_model}.

\begin{figure*}
\gridline{\fig{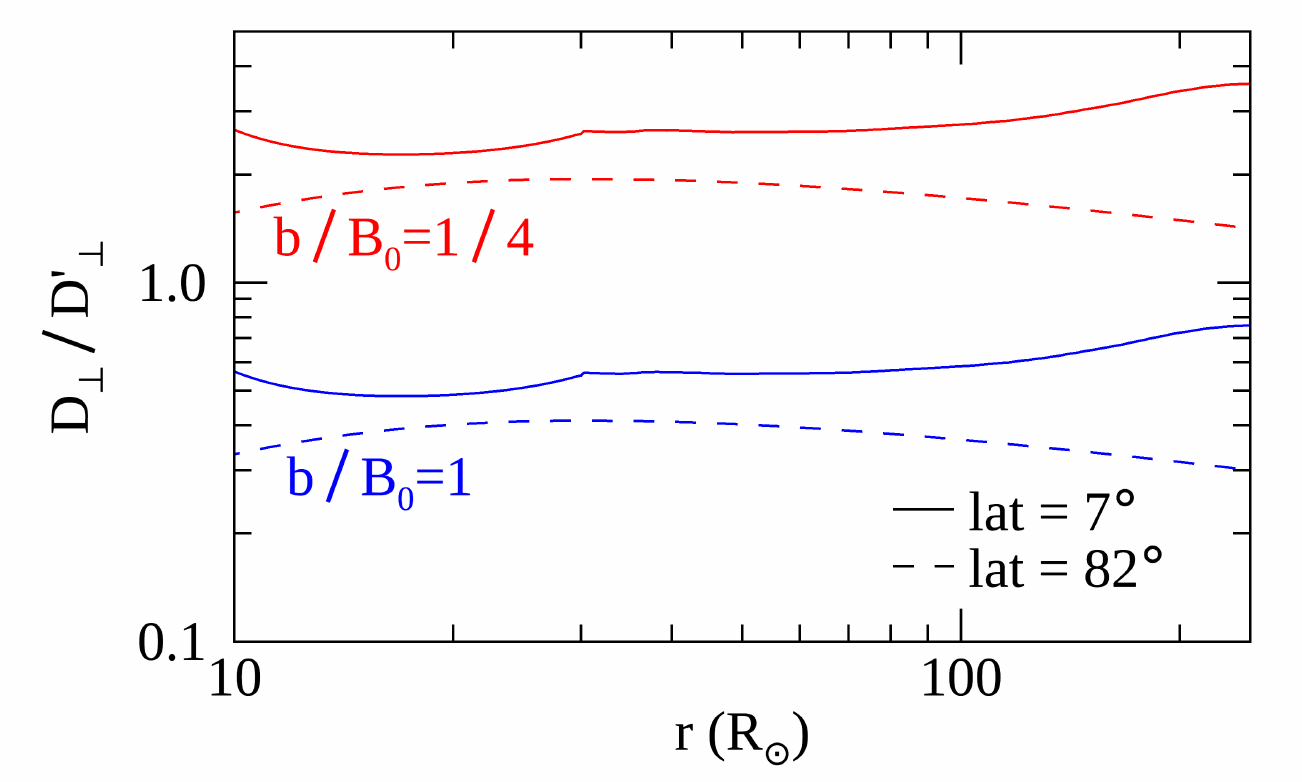}{.5\textwidth}{(a)}
         \fig{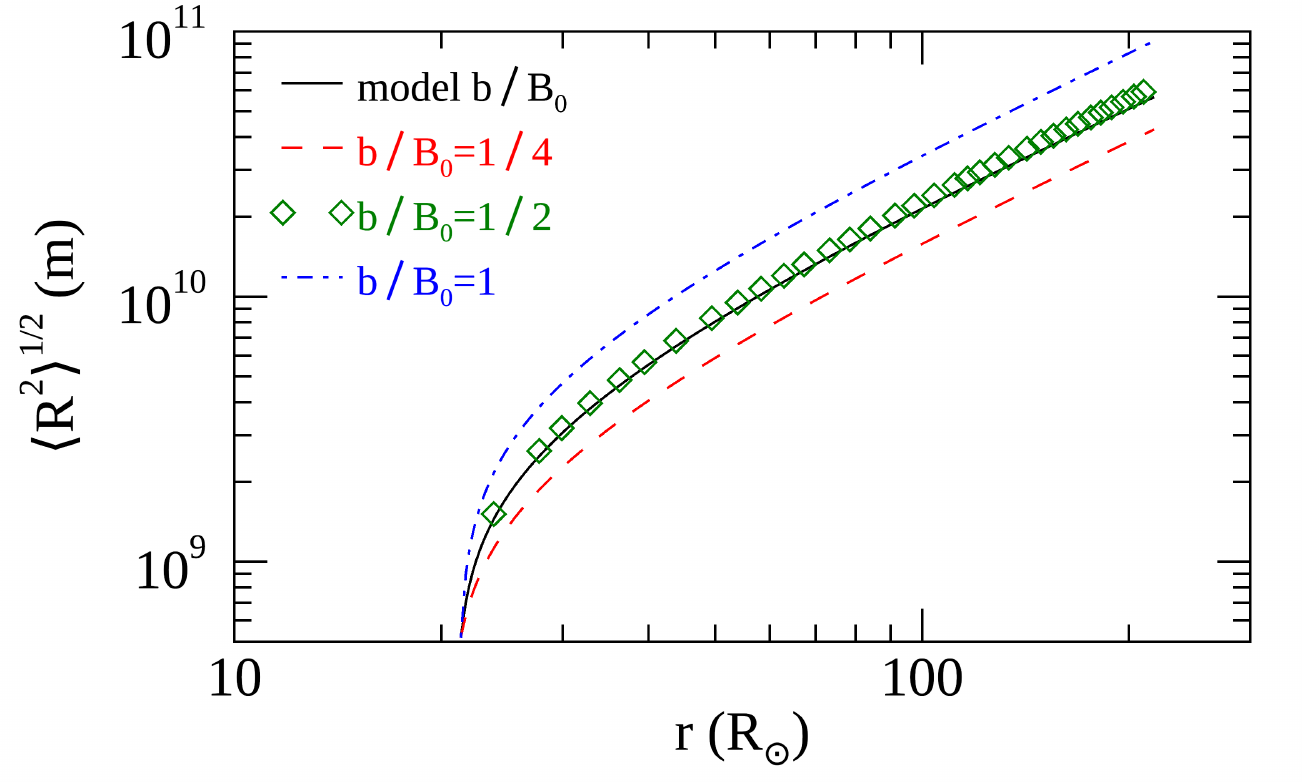}{.5\textwidth}{(b)}
         }
\caption{(a) Ratio of FLRW diffusion coefficients \(D_\perp/D'_\perp\), where \(D_\perp\) is computed from the turbulence transport model and \(D'_\perp\) using the simple approximations described in Appendix B. The figure shows the radial evolution of this ratio for two field lines at latitudes of 7\degree\ and 82\degree, at a longitude of 0\degree, from the tilted-dipole simulation (Run I; Section \ref{sec:dipole}). (b) Rms spread of a selected field line (0\degree~longitude and 55\degree~latitude) Run I. The solid black curve shows the calculation based on the turbulence transport model, while the three other curves are obtained using a constant \(b/B_0\) and the \citet{Hollweg1986JGR} prescription for the correlation scale. Note that the \(b/B_0=1/2\) case is not shown in panel (a) to avoid cluttering the figure.}
\label{fig:app_simple}
\end{figure*}

Of the three values for the ratio \(b/B_0\) considered here, 1 is more consistent with our turbulence transport model as well as with observations in the inner heliosphere \citep{bruno2013LRSP}. However, the Hollweg prescription for \(\lambda\) is, in general, larger than the correlation scale from our turbulence transport model. Therefore, \(b/B_0 = 1\) yields a diffusion coefficient [\(D'_\perp\) in Figure 12(a)] 
that is larger than the one obtained from the turbulence transport model, and a correspondingly larger rms spread [Figure 12(b)]. 
A value of \(b/B_0=1/2\) matches the rms spread from the turbulence model rather well, especially for heliocentric distances smaller than \(\sim 100~\rs\).

\bibliography{chhibref}

\end{document}